\documentclass[twocolumn, aps, 10pt, floatfix]{revtex4-1}

\usepackage{lmodern}
\usepackage[T1]{fontenc}
\usepackage[utf8]{inputenc}
\usepackage[english]{babel}

\usepackage{amsmath}
\usepackage{amssymb}
\usepackage{mathtools}
\usepackage{bm}
\usepackage{graphicx,color}
\usepackage{hyperref}
\usepackage{booktabs}
\usepackage{epstopdf}

\mathtoolsset{showonlyrefs}

\hypersetup{
  colorlinks   = true, 
  urlcolor     = blue, 
  linkcolor    = blue, 
  citecolor   = red, 
}

\newcommand{\ic}{\ensuremath{\mathrm{i}}}
\newcommand{\dee}{\ensuremath{\mathrm{d}}}
\newcommand{\ket}[1]{|#1\rangle}
\newcommand{\bra}[1]{\langle#1|}

\def\Henkel{henkel_1999}
\def\diFrancesco{francesco_2012}
\def\CFTped{belavin_infinite_1984, ginsparg_applied_1988, \diFrancesco, \Henkel}

\def\eye{I}
\def\CFT{\textsl{\tiny CFT}}
\def\TFI{\textsl{\tiny Ising}}
\def\potts{\textsl{\tiny Potts}}
\def\ANNNI{\textsl{\tiny ANNNI}}

\begin{document}

\title{Extraction of conformal data in critical quantum spin chains \\ using the Koo-Saleur formula}


\author{Ashley Milsted}
\email[]{amilsted@pitp.ca}
\affiliation{Perimeter Institute for Theoretical Physics, Waterloo ON, N2L 2Y5, Canada}
\author{Guifré Vidal}
\affiliation{Perimeter Institute for Theoretical Physics, Waterloo ON, N2L 2Y5, Canada}

\date{\today}

\begin{abstract}
  We study the emergence of two-dimensional conformal symmetry in critical quantum spin chains on the finite circle. Our goal is to characterize the conformal field theory (CFT) describing the universality class of the corresponding quantum phase transition. As a means to this end, we propose and demonstrate automated procedures which, using only the lattice Hamiltonian \mbox{$H = \sum_j h_j$} as an input, systematically identify the low-energy eigenstates corresponding to Virasoro primary and quasiprimary operators, and assign the remaining low-energy eigenstates to conformal towers. The energies and momenta of the primary operator states are needed to determine the primary operator scaling dimensions and conformal spins -- an essential part of the conformal data that specifies the CFT. Our techniques use the action, on the low-energy eigenstates of $H$, of the Fourier modes $H_n$ of the Hamiltonian density $h_j$. The $H_n$ were introduced as lattice representations of the Virasoro generators by Koo and Saleur [Nucl. Phys. B 426, 459 (1994)]. In this paper we demonstrate that these operators can be used to extract conformal data in a \textit{nonintegrable} quantum spin chain. 
\end{abstract}

\maketitle

\section{Introduction}
\label{sec:introduction}

Conformal field theory (CFT) \cite{\CFTped} is ubiquitous in modern theoretical physics. It describes fixed points of the renormalization group flow \cite{cardy_scaling_1996}, making it central to our understanding of quantum field theory \cite{wilson_renormalization_1974}. It is also a core component both of string theory \cite{tong_lectures_2009} and of the AdS/CFT correspondence of quantum gravity \cite{maldacena_1999}. In condensed matter, as well as in statistical mechanics, continuous phase transitions can often be understood in terms of an underlying CFT that describes their universal, long-distance/low-energy physics \cite{belavin_infinite_1984, cardy_scaling_1996, \diFrancesco, \Henkel}. 
Based on a previous proposal by Koo and Saleur \cite{koo_representations_1994}, in this paper we develop tools to investigate the emergence of conformal symmetry in generic quantum spin chains at criticality.

In order to present our results, we first need to recall two well-known facts about CFTs in two spacetime dimensions \cite{\CFTped}. (i) On the plane, parameterized by a complex coordinate $z=x+iy$, a CFT contains infinitely many scaling operators $\varphi_\alpha(z)$. These are fields that transform covariantly under a rescaling of the plane by a factor $\lambda > 0$ or a rotation by an angle $\theta \in [0,2\pi)$:
\begin{equation} \label{eq:CFT_fields}
\begin{alignedat}{2}
z&\rightarrow  \lambda z~~ \mbox{(rescaling)}\quad&\Leftrightarrow\quad  \varphi_\alpha(0) &\rightarrow \lambda^{-\Delta_\alpha} \;\varphi_\alpha(0), \\
z& \rightarrow e^{\ic \theta} z~~ \mbox{(rotation)}\quad&\Leftrightarrow\quad  \varphi_\alpha(0) &\rightarrow e^{-\ic \theta S_\alpha} \;\varphi_\alpha(0),
\end{alignedat}
\end{equation}
where $\Delta_\alpha$ and $S_\alpha$ are the \emph{scaling dimension} and \emph{conformal spin} of $\varphi_\alpha(z)$. Scaling operators are organized into \textit{conformal towers}, each consisting of a Virasoro \textit{primary} operator and its \textit{descendants} (see, e.g., Fig.~\ref{fig:ising_towers}). (ii) The operator-state correspondence establishes that for each scaling operator $\varphi_\alpha$ there is an eigenstate $\ket{\varphi_\alpha}$ of the CFT Hamiltonian $H^{\CFT}$ on the circle, with energy and momentum given 
by
\begin{align} \label{eq:CFTspec}
\begin{split}
  E^\CFT_\alpha = \frac{2\pi}{L}\left( \Delta_\alpha - \frac{c}{12} \right), \qquad
  P^\CFT_\alpha = \frac{2\pi}{L} S_\alpha,
\end{split}
\end{align}
where $L$ is the length of the circle and $c$ is the \textit{central charge} of the CFT, which determines the Casimir energy. The scaling dimensions, conformal spins and \textit{operator product expansion} (OPE) coefficients (three-point correlators) of the primary operators, together with the central charge, fully characterize the CFT \cite{belavin_infinite_1984} and are referred to as \emph{conformal data}.

\subsection{Extraction of conformal data}
The extraction of conformal data from lattice models has a long history. Following the landmark 1984 publication by Belavin, Polyakov and Zamolodchikov of \cite{belavin_infinite_1984}, which revealed the intricate structure of 2D CFT, Cardy, Blöte, Nightingale and Affleck \cite{cardy_conformal_1984, blote_1986, cardy_operator_1986, cardy_logarithmic_1986, affleck_universal_1986} discovered that, at low energies and after suitably normalizing the lattice Hamiltonian $H$, the energies and momenta of a critical quantum spin chain made of $N$ spins must read
\begin{align} \label{eq:FS_spec}
  E_\alpha = \frac{2\pi}{N}(\Delta_\alpha - \frac{c}{12}) + \mathcal{O}(N^{-x}), \quad P_\alpha = \frac{2\pi}{N} S_\alpha.
\end{align}
This matches the CFT spectrum~\eqref{eq:CFTspec} up to subleading, non-universal corrections $\mathcal{O}(N^{-x})$, where $x > 1$ is also model-specific 
\footnote{In models with marginal operators, there may also be logarithmic finite-size corrections \cite{cardy_logarithmic_1986}. We have successfully tested the proposals of this paper also on such models, including the four-state Potts quantum spin chain.}. One can therefore estimate the scaling dimensions $\Delta_{\alpha}$ and conformal spins $S_{\alpha}$ of the CFT from the energies $E_{\alpha}$ and momenta $P_{\alpha}$ computed on the lattice, see e.g.\ Fig.~\ref{fig:ising_lattice_spec}. This result has proved extremely useful in understanding critical lattice systems e.g.~\cite{gehlen_conformal_1986, gehlen_operator_1986, henkel_finite-size_1987, alcaraz_1987a, baake_operator_1987, balbao_operator_1987, allton_1+12_1988, affleck_critical_1989, alcaraz_xxz_1989, frahm_critical_1990, voit_one-dimensional_1995, boschi_mathsfc_2003, feiguin_interacting_2007, fuehringer_dmrg_2008, xavier_entanglement_2010, gendiar_suppression_2011, katsura_sine-square_2012, wen_evolution_2016}.

One can think of \eqref{eq:FS_spec} as demonstrating a low-energy correspondence between the critical lattice Hamiltonian $H$ and the CFT Hamiltonian~$H^{\CFT}$
\begin{align} \label{eq:HsimHCFT}
~~~~H = \sum_{j=1}^N h_j \quad \sim \quad H^\CFT = \int_0^L \dee x \; h^\CFT(x),
\end{align}
where $h_j$ and $h^\CFT(x)$ denote the lattice and continuum Hamiltonian densities. It is then natural to ask whether this \emph{global} correspondence extends to the \emph{local} densities
\begin{align} \label{eq:hsimhCFT}
  h_j ~~ \sim ~~ h^\CFT(x).
\end{align}
An example of this local correspondence was already found in 1971 by Kadanoff and Ceva \cite{kadanoff_determination_1971}, who showed that a lattice analogue of the energy-momentum tensor exists in the Ising model. Later, Koo and Saleur \cite{koo_representations_1994} demonstrated the principle more generally by showing that, in some integrable models, the Fourier modes $H_n$ of $h_j$, defined so that \mbox{$h_j = (2\pi/N^2) \sum_{n} e^{-\ic j n \frac{2\pi}{N}} H_n$}, behave as lattice representations of certain linear combinations of the Virasoro generators of conformal symmetry: $H_n^{\CFT} \equiv L^\CFT_n + \overline{L}{}^{\CFT}_{-n} - \delta_{n,0} c/12$.
Other work had previously established the existence of lattice representations (or \emph{deformations}) of (parts of) the Virasoro algebra in certain integrable systems \cite{itoyama_lattice_1987, volkov_miura_1988, faddeev_liouville_1988, babelon_exchange_1990, johannesson_lattice_1992, volkov_quantum_1992, belov_q-deformation_1993, kellendonk_virasoro_1993}, but the proposal of \cite{koo_representations_1994} is of particular importance because it provides a prescription for constructing lattice analogues of \emph{all} the Virasoro generators. This provides access to a wealth of information about the CFT, including the central charge. Indeed, a number of authors have used the so-called Koo-Saleur formula to extract conformal data in various models, especially \emph{logarithmic} CFTs, which are nontrivial nonunitary CFTs with $c=0$, see e.g.~\cite{read_associative-algebraic_2007, dubail_conformal_2010, vasseur_puzzle_2012, gainutdinov_logarithmic_2013, gainutdinov_lattice_2013, bondesan_chiral_2015}. However, as yet the Koo-Saleur formula has not enjoyed the same widespread use as \eqref{eq:FS_spec}, having been applied only to integrable systems.

\subsection{Our results}

In this paper we propose and test methods which apply the Hamiltonian-density Fourier modes $H_n$ \cite{koo_representations_1994} to systematically identify low-energy eigenstates of a critical spin chain Hamiltonian $H$ (with local interactions) with CFT scaling operators. In particular, we present automated procedures for finding the eigenstates corresponding to primary and quasiprimary operators, as well as for assigning all remaining low-energy eigenstates to their respective (Virasoro or global) conformal towers. A key feature of these methods is that they provide a general means for determining which scaling dimensions and conformal spins derived from \eqref{eq:FS_spec} belong to primary fields in the CFT, thus delivering a crucial piece of the conformal data. They also deliver an improved way of identifying the energy-momentum-tensor state, often used to determine the correct normalization for $H$. Furthermore, our construction sets the stage for a systematic determination of the OPE coefficients for generic critical spin chains, which involves additionally determining scaling operators on the lattice and will be discussed in~\cite{zou_upcoming_2017}.

Finally, we establish that our methods, and hence the Koo-Saleur formula, are applicable away from integrability by demonstrating them in the self-dual ANNNI model: a nonintegrable perturbation of the Ising model.

We stress that, although for this paper we used exact diagonalization to obtain the low-energy eigenstates of $H$, our core proposal is \emph{independent} of the method used to obtain these eigenstates. Indeed, we can also apply operators $H_n$ to energy eigenstates obtained with more sophisticated techniques, such as periodic \textit{matrix product states} \cite{zou_conformal_2017}, and in this way analyze larger systems, which carry smaller finite-size errors.

Note: Throughout the paper we differentiate between lattice objects, such as $H$, $P$, and $H_n$, and their CFT counterparts $H^{\CFT}$, $P^{\CFT}$, and $H_n^{\CFT}$, by means of the superscript $^\CFT$. On the other hand, states denoted as $\ket{\varphi}$, $\ket{\varphi_{\alpha}}$, etc.\ belong either to the lattice or the CFT, as can be determined from the context. 

\section{Low-energy correspondence for Hamiltonian densities}
\label{sec:LowEnergy}

\subsection{Critical quantum spin chains and CFTs}
We consider a periodic 1D lattice made of $N$ sites with a translation invariant quantum Hamiltonian  
\begin{align} \label{eq:Hlat}
  H = \sum_{j=1}^N h_j,
\end{align}
that decomposes as a sum of local Hamiltonian terms, where the term $h_j$ is located about site $j$ and will be referred to as the Hamiltonian density on that site. A canonical example is the transverse field Ising model
\begin{align} \label{eq:H_TFI}
  H^\TFI(\lambda) \equiv - \sum_{j=1}^N \left[ \sigma^X_j \sigma^X_{j+1} + \lambda\sigma^Z_j \right],
\end{align}
which is critical at $\lambda = 1$. We assume that, at criticality, there is a corresponding quantum CFT Hamiltonian
\begin{align} \label{eq:HCFT}
 H^\CFT =  \int_0^L \dee x \; h^\CFT(x),
\end{align}
where $x \in (0, L]$ parameterizes a circle of radius $L/2\pi$ and the Hamiltonian-density field operator $h^\CFT(x)$ can be written \cite{\CFTped} in terms of the chiral and anti-chiral components $T^\CFT(x)$ and $\overline{T}{}^\CFT(x)$ of the traceless energy-momentum tensor of the CFT on the circle,
\begin{align} \label{eq:hCFT_T}
  h^\CFT(x) \equiv \frac{1}{2\pi}\left(T^\CFT(x) + \overline{T}{}^\CFT(x)\right).  
\end{align}
Similarly, to the lattice momentum operator $P$ (defined such that $e^{\ic P \frac{2\pi}{N}}$ is a translation by one lattice site) we associate the CFT momentum operator
\begin{align} \label{eq:PCFT}
 P^\CFT =  \int_0^L \dee x \; p^\CFT(x),
\end{align} 
where $p^\CFT(x) \equiv \left(T^\CFT(x) - \overline{T}^\CFT(x)\right)/(2\pi)$ is the momentum density. 

\subsection{Fourier mode expansions}

The Fourier modes $L{}^\CFT_n$ and $\overline{L}{}^\CFT_n$ of the chiral and anti-chiral energy-momentum tensor operators \cite{\CFTped}
\begin{align} \label{eq:LCFT} 
  \begin{split}
  L^\CFT_n &\equiv \frac{L}{(2\pi)^2} \int_0^{L} \dee x \; e^{+\ic n x \frac{2\pi}{L}} \, T^\CFT(x) + \frac{c}{24}\delta_{n,0}, \\
  \overline{L}^\CFT_n &\equiv \frac{L}{(2\pi)^2} \int_0^{L} \dee x \; e^{-\ic n x \frac{2\pi}{L}} \, \overline{T}^\CFT(x) + \frac{c}{24}\delta_{n,0},
  \end{split}
\end{align}
where $n\in \mathbb{Z}$, furnish chiral and anti-chiral instances of the Virasoro algebra \cite{virasoro_subsidiary_1970, belavin_infinite_1984}
\begin{align} \label{eq:Virasoro} 
  \begin{split}
  [ L^\CFT_n, L^\CFT_m ] &= (n-m)L^\CFT_{n+m} + \frac{c}{12} n(n^2-1)\delta_{n+m,0} \\
  [ L^\CFT_n, \overline{L}^\CFT_m ] &= 0 \\
  [ \overline{L}^\CFT_n, \overline{L}^\CFT_m ] &= (n-m)\overline{L}^\CFT_{n+m} + \frac{c}{12} n(n^2-1)\delta_{n+m,0}
  \end{split}
\end{align}
and are the canonical choice of generators of conformal symmetry on the CFT Hilbert space. 

Importantly for our purposes, the Fourier modes $H^\CFT_n$ of the Hamiltonian density operator $h^\CFT(x)$ correspond to certain linear combinations of the above Virasoro generators, 
\begin{align} \label{eq:Hn_L_CFT} 
  H^\CFT_n &\equiv \frac{L}{2\pi} \int_0^{L} \dee x \; e^{+\ic n x \frac{2\pi}{L}} \, h^\CFT(x)\\&=L^\CFT_n + \overline{L}^\CFT_{-n} - \frac{c}{12}\delta_{n,0}.
\end{align}
where we note that, for $n=0$
\begin{equation}
H_0^{\CFT} = L_0^{\CFT} + \overline{L}{}^{\CFT}_0 - \frac{c}{12} = \frac{L}{2\pi} H^{\CFT}.
\end{equation}
 
In direct analogy, following the proposal of Koo and Saluer \cite{koo_representations_1994}, we introduce the Fourier modes $H_n$ of the lattice Hamiltonian density $h_j$
\begin{align} \label{eq:Hn} 
  H_n \equiv \frac{N}{2\pi} \sum_{j=1}^N e^{+\ic j n \frac{2\pi}{N}} h_j,~~~~~~ H_0 = \frac{N}{2\pi} H,
\end{align}
in terms of which the lattice Hamiltonian density $h_j$ at site $j$ reads
\begin{align}
  h_j = \frac{2\pi}{N^2} \sum_{n=-N/2}^{+N/2} e^{-\ic j n \frac{2\pi}{N}} H_n.
\end{align}

\subsection{General strategy}

Our goal is to use the Fourier modes $H_n$ of the lattice Hamiltonian density $h_j$ to systematically extract conformal data from the low-energy subspace of the critical lattice Hamiltonian $H$. This will be discussed in Sect.~\ref{sec:extraction} and then numerically demonstrated in Sect.~\ref{sec:results}.

The central assumption of is that, at low energies and up to finite-size corrections, each $H_n$ should act on the simultaneous eigenstates $\ket{\varphi_\alpha}$ of $H$ and $P$ on the lattice as its CFT counterpart $H_n^{\CFT}$ does on the simultaneous eigenstates of $H^{\CFT}$ and $P^{\CFT}$ in the continuum. Strong evidence for this was provided in \cite{koo_representations_1994} and subsequent work (for integrable systems), but we will need more details for our purposes.
We therefore begin in Sect.~\ref{sec:ConformalTowers} by explaining how the Fourier modes $H_n^{\CFT}$ act in the continuum. This is best understood in terms of the Fourier modes $L{}^\CFT_n$ and $\overline{L}{}^\CFT_n$, which act simply as ladder operators on the eigenbasis $\ket{\varphi_\alpha}$.

At this point, a natural question to ask is whether it would be more convenient to construct, and directly work with, lattice versions $L_n$ and $\overline{L}_n$ of the Virasoro generators ${}L^\CFT_n$ and $\overline{L}{}^\CFT_n$, as was done in \cite{koo_representations_1994}, instead of using the lattice Fourier modes $H_n$. After all, most CFT practitioners are already familiar with the Virasoro generators $L{}^\CFT_n$ and $\overline{L}{}^\CFT_n$, which explicitly discriminate between chiral and anti-chiral CFT modes, and not so much with the Fourier modes $H^\CFT_n$. As explained in App.~\ref{app:mom}, doing so is possible in principle but far from optimal in practice. Next we briefly summarize why. 

Given the lattice Hamiltonian density $h_j$ as the only input, it is indeed possible to use energy conservation to obtain a lattice momentum density $p_j \equiv i \left[h_j, h_{j-1} \right]$, and thus produce chiral and anti-chiral energy-momentum operators $T_j = \frac{1}{2}(h_j + p_j)$ and $\overline{T}_j = \frac{1}{2}(h_j - p_j)$, whose Fourier mode expansion leads to lattice Virasoro generators $L_n$ and $\overline{L}_n$ that act as $L{}^\CFT_n$ and $\overline{L}{}^\CFT_n$ at low energies and up to finite size corrections. However, by construction there are additional finite-size corrections in $L_n$ and $\overline{L}_n$, compared to $H_n$, which can be traced back to finite-size corrections to the eigenstate energies of $H$ (see App.~\ref{app:mom}). Therefore, from a numerical perspective, it is preferable to work with the lattice Fourier modes $H_n$, as we do in this paper.

\section{Conformal towers in the continuum}
\label{sec:ConformalTowers}

\subsection{The Virasoro generators as ladder operators}
\label{subsec:CFTstruct}

\begin{figure}
\includegraphics[width=\linewidth]{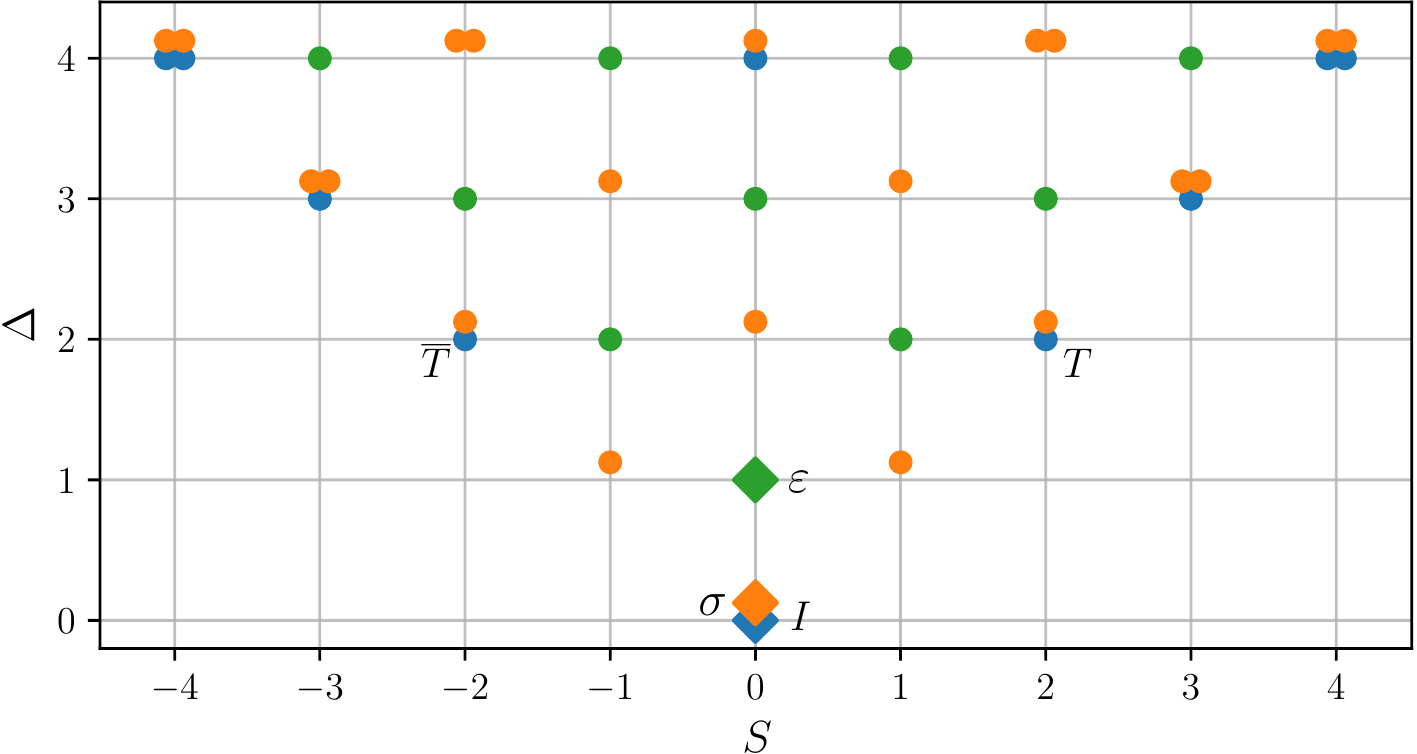}
\caption{\label{fig:ising_towers} Exact spectrum of the Ising CFT Hamiltonian in terms of $\Delta$ and $s$, color-coded by conformal tower, showing the location of the primary states $|\eye\rangle$, $|\sigma\rangle$ and $|\varepsilon\rangle$, and the energy-momentum states $|T\rangle$ and $|\overline{T}\rangle$. {\bf Note}: We shift points horizontally from their allowed values ($S$ is quantized) to avoid overlaps and better show degeneracies in this and subsequent figures.
}
\end{figure}

Recall that in a 2D CFT, the combinations $L^\CFT_0 \pm \overline{L}{}^\CFT_0$ generate the dilations and rotations in~\eqref{eq:CFT_fields} \cite{\CFTped}. Therefore, by the operator-state correspondence \cite{fubini_new_1973, belavin_infinite_1984}, these operators act on the state $\ket{\varphi_\alpha}$ as
\begin{align}
\left(L^\CFT_0 + \overline{L}{}^\CFT_0\right) \ket{\varphi_\alpha} &= \Delta_{\alpha}\ket{\varphi_\alpha},\\
\left(L^\CFT_0 - \overline{L}{}^\CFT_0\right) \ket{\varphi_\alpha} &= S_{\alpha} \ket{\varphi_\alpha},
\end{align}
which, given that  $H^\CFT$ and $P^\CFT$ can be written in terms of $L^\CFT_0 \pm \overline{L}{}^\CFT_0$ as
\begin{align} \label{eq:HPCFT_L}
\begin{split}
  H^\CFT &= \frac{2\pi}{L}\left( L^\CFT_0 + \overline{L}^\CFT_0 - \frac{c}{12}\right) \\
  P^\CFT &= \frac{2\pi}{L}\left( L^\CFT_0 - \overline{L}^\CFT_0 \right),
\end{split}
\end{align}
automatically implies \eqref{eq:CFTspec} or, equivalently,
\begin{align} \label{eq:CFTspec_2}
\begin{split}
  \Delta_\alpha = \frac{L}{2\pi} E^\CFT_\alpha + \frac{c}{12}, \qquad
  S_\alpha = \frac{L}{2\pi} P^\CFT_\alpha.
\end{split}
\end{align}

\begin{figure}
\includegraphics[width=\linewidth]{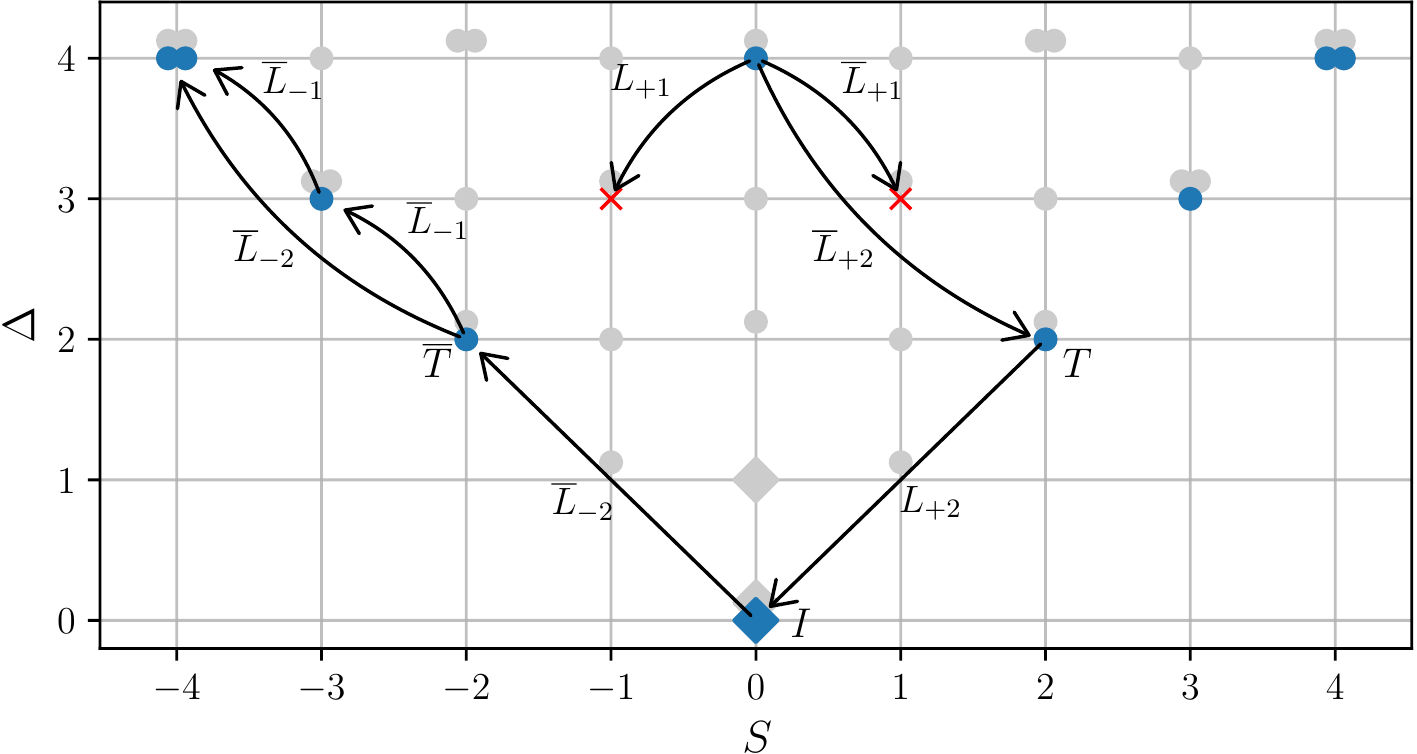}
\caption{\label{fig:ising_ladders} Illustration of the action of the ladder operators (Virasoro generators) on the energy eigenstates of the Ising CFT Hamiltonian belonging to the $\eye$ conformal tower. Two possible paths from $(\Delta\!=\!4,S\!=\!0)$ to $(\Delta\!=\!4,S\!=\!-4)$ are shown, as is the annihilation of the quasiprimary state \mbox{$|\Delta\!=\!4,S\!=\!0\rangle$} by $\overline{L}_{+1}$ and $L_{+1}$.
}
\end{figure}

Let us temporarily denote $\ket{\varphi_\alpha}$ as $\ket{\Delta_\alpha,S_\alpha}$. From \eqref{eq:HPCFT_L} and the Virasoro algebra \eqref{eq:Virasoro} it can be seen that the Virasoro generators are \emph{ladder operators} of $H^\CFT$ and $P^\CFT$. They indeed act on an eigenstate $|\Delta_\alpha, S_\alpha\rangle$ as
\begin{align} \label{eq:Ladders}
\begin{split}
  L^\CFT_{n}|\Delta_\alpha, S_\alpha\rangle &\propto |\Delta_\alpha \!- n,\; S_\alpha \!- n\rangle, \\
  \overline{L}{}^\CFT_{n}|\Delta_\alpha, S_\alpha\rangle &\propto |\Delta_\alpha \!- n,\; S_\alpha \!+ n\rangle,
\end{split}
\end{align}
raising $\Delta$ for $n<0$ and lowering it for $n>0$. Note also that $L^\CFT_{n}$ and $\overline{L}{}^\CFT_{n}$ change $S$ in opposite directions. This is illustrated in Fig.~\ref{fig:ising_ladders}.

The Virasoro operators $L^\CFT_n$, $\overline{L}{}^\CFT_n$ generate multiple (generally an infinite number, but not always \cite{friedan_conformal_1984}) distinct towers of eigenstates of $H^\CFT$ and $P^{\CFT}$ called \emph{conformal towers}. Each tower has, at its base, a \emph{primary state} (corresponding to a \emph{primary operator}). Primary states are therefore those states annihilated by all ladder operators that reduce the energy \cite{\CFTped}: $L^\CFT_{n}|\varphi\rangle = \overline{L}{}^\CFT_{n}|\varphi\rangle = 0$ for all $n>0$. By \eqref{eq:Virasoro}, $L^\CFT_{-1}, L^\CFT_{-2}$ generate the subalgebra $L^\CFT_n$ for $n<0$ (and similar for $\overline{L}{}^\CFT_n$) so that this condition is equivalent to:
\begin{align} \label{eq:Pstates}
  |\varphi\rangle \text{ primary } \Leftrightarrow ~L^\CFT_{n} |\varphi\rangle = \overline{L}{}^\CFT_n |\varphi\rangle = 0,~~ n=1,2.~~~~
\end{align}
By acting with products of powers of $L^\CFT_{n}$, $\overline{L}{}^\CFT_{n}$ with $n<0$ on a primary $|\varphi\rangle$, all \emph{descendant states} in its tower can be reached. From \eqref{eq:Ladders}, descendants $|\varphi'\rangle$ of a primary $|\varphi\rangle$ must have
scaling dimension $\Delta_{\varphi'}$ and conformal spin $S_{\varphi'}$ given by
\begin{equation}
\label{eq:possibleDS}
\Delta_{\varphi'} = \Delta_\varphi + n, ~~~~ 
S_{\varphi'} = S_\varphi \pm m, 
    \quad \text{for } n \geq m, ~~ 
\end{equation} 
where $n \in \mathbb{N}$ and $m \in \mathbb{Z}$. Furthermore, it follows from \eqref{eq:Pstates} that all descendants can be reached from the primary using only  $L^\CFT_{-n}$, $\overline{L}{}^\CFT_{-n}$ with $n=1,2$.
 
Let us pause here and briefly consider a simple example to which we will return later: The \emph{Ising CFT} only has three primary operators \cite{friedan_conformal_1984}:
\begin{center}
\renewcommand{\arraystretch}{1.2} 
\begin{tabular}{ @{}ccccc@{} }
  \toprule[0.8pt]
    primary operator & $\;\Delta\;$ & $\;S\;$ & state \\ \midrule[0.4pt]
  identity  $\eye$ & $0$ & $0$ & $|\eye\rangle$ \\
  spin  $\sigma(x)$ & $1/8$ & $0$ & $|\sigma\rangle$ \\
  energy density  $\varepsilon(x)$ ~~& $1$ & $0$ & $|\varepsilon\rangle$ \\
  \bottomrule[0.8pt]
\end{tabular}
\end{center}
Therefore it has just three conformal towers. From this data we can infer information about the spectrum of $H^\CFT$, $P^\CFT$ using \eqref{eq:CFTspec} and \eqref{eq:possibleDS}. For example, all eigenstates have either $\Delta_\alpha \in \mathbb{N}$ (descendants of $|\eye\rangle$ and $|\varepsilon\rangle$) or $\Delta_\alpha \in \mathbb{N} + \frac{1}{8}$ (descendants of $|\sigma\rangle$). The low-energy spectrum of the Ising CFT is shown in Fig.~\ref{fig:ising_towers}. In Fig.~\ref{fig:ising_ladders} we illustrate how the ladder operators can be used to connect states within a particular conformal tower.

\subsection{Identity, energy-momentum, and central charge}
\label{subsec:Identity}

Returning to a generic 2D CFT, a particularly important primary state that is always present is the ``identity state'' $|\eye\rangle$. In a unitary CFT, which is the main focus of this work, the state $|\eye\rangle$ corresponds to the ground state of the Hamiltonian $H^\CFT$. This state is unique in having a vanishing scaling dimension $\Delta_\eye = 0$ and in being annihilated by all $L^\CFT_{n}, \overline{L}{}^\CFT_{n}$ with $n=0,\pm 1$, which are the generators of \emph{global} conformal transformations (those that are well-defined throughout the 2D plane) \cite{\CFTped}. 

Another relevant notion is that of a \emph{quasiprimary} state~\cite{\CFTped}, defined as a state that is annihilated by both $L^\CFT_{1}$ and $\overline{L}{}^\CFT_{1}$:
\begin{equation} 
  |\varphi\rangle \text{ quasiprimary} ~~\Leftrightarrow ~~L{}^\CFT_{1} |\varphi\rangle = \overline{L}{}^\CFT_{1} |\varphi\rangle = 0.\label{eq:QPstates}
\end{equation}
This includes all primary states, but also certain descendant states. Two important quasiprimary states that are present in any CFT are those corresponding to the CFT energy-momentum operators $T{}^\CFT(x)$ and $\overline{T}{}^\CFT(x)$. They are descended from the ground state $|\eye\rangle$ as
\begin{align} \label{eq:TfromI}
  \sqrt{\frac{c}{2}}|T\rangle = L{}^\CFT_{-2}|\eye\rangle \:\text{ and }\: \sqrt{\frac{c}{2}}|\overline{T}\rangle = \overline{L}{}^\CFT_{-2}|\eye\rangle,
\end{align}
where $c$ is the central charge, and thus have scaling dimensions \mbox{$\Delta_T=\Delta_{\overline{T}}=2$} and conformal spins $S_T = 2$, $S_{\overline{T}} = -2$. For the Ising CFT, states $\ket{\eye}$, $\ket{T}$, and $\ket{\overline{T}}$ can be seen in Figs. \ref{fig:ising_towers} and \ref{fig:ising_ladders}.

\subsection{Characterization in terms of $H_n$}

Finally, we have to translate the above statements for the Virasoro generators $L^\CFT_{n}$, $\overline{L}{}^\CFT_{n}$ into statements for the Fourier modes $H_n^{\CFT}$ of the Hamiltonian density defined in \eqref{eq:Hn_L_CFT}. Recalling that the Fourier modes $H_n$ for $n\not = 0$ are linear combinations of the Virasoro generators, \mbox{$H_{n}^\CFT = L{}^\CFT_n + \overline{L}{}^\CFT_{-n}$}, we can infer their behavior from~\eqref{eq:Ladders}:
\begin{equation} \label{eq:HnCFT_action}
  \begin{aligned}
  H^\CFT_{n} \;|\Delta_\alpha, S_\alpha\rangle =\: &a\;|\Delta_\alpha \!- n,\; S_\alpha \!- n\rangle \;+ \\
 &b\;|\Delta_\alpha \!+ n,\; S_\alpha \!- n\rangle,
  \end{aligned}
\end{equation}
where $a$ and $b$ are determined by conformal symmetry and may equal zero~\cite{\CFTped}. The following simple observation will also prove very useful. Given an energy eigenstate $\ket{\varphi}$ with energy $E_{\varphi}$, let $\Gamma_{\varphi}$ be a projector onto all the eigenstates with energy smaller than $E_{\varphi}$,
\begin{equation}
 \Gamma_{\varphi} \equiv \sum_{\varphi_\alpha: E_\alpha < E_{\varphi}} \ket{\varphi_\alpha}\bra{\varphi_\alpha}.
\end{equation}
Then we have that the product $\Gamma_{\varphi} \, H_n^{\CFT}$ acts on $\ket{\varphi}$ as would either just $L_n^{\CFT}$ or $\overline{L}{}^{\CFT}_{-n}$ according to
\begin{align} \label{eq:HnLn}
 \Gamma_{\varphi} \; H^\CFT_{n} \; |\varphi\rangle = \begin{cases}
   L^\CFT_{n} \;|\varphi\rangle & \text{if } n < 0, \\
   \overline{L}{}^\CFT_{-n} \;|\varphi\rangle & \text{if } n > 0.
 \end{cases}
\end{align}

It follows that we can recast the characterization~\eqref{eq:Pstates} of a primary state as
\begin{align} \label{eq:PstatesHn}
  |\varphi\rangle \text{ primary } \Leftrightarrow ~ \Gamma_{\varphi} \; H^\CFT_{n} |\varphi\rangle = 0,~~ n=\pm 1,\pm 2,~~~~~
\end{align}

Similarly, the characterization~\eqref{eq:QPstates} of a quasiprimary state reads 
\begin{align} \label{eq:QPstatesHn}
  |\varphi\rangle \text{ quasiprimary } \Leftrightarrow ~ \Gamma_{\varphi} \; H^\CFT_{n} |\varphi\rangle = 0,~~ n=\pm 1.~~~~~
\end{align}
More generally, by using either Eq.~\eqref{eq:HnLn} or a similar expression with a complementary projector $\mathbb{I}-\Gamma_\varphi$, we can use the Fourier modes $H_n$ of the Hamiltonian density $h(x)$ to reproduce the action of the Virasoro generators $L_n^{\CFT}$ and $\overline{L}{}^{\CFT}_n$. 
Finally, an expression such as \eqref{eq:TfromI} translates directly into
\begin{align} \label{eq:TfromIH2}
  \sqrt{\frac{c}{2}}|T\rangle = H^\CFT_{-2}|\eye\rangle \:\text{ and }\: \sqrt{\frac{c}{2}}|\overline{T}\rangle = H^\CFT_{2}|\eye\rangle,
\end{align}
without the need of projectors, given that there are no states with energy below that of $\ket{\eye}$.

\section{Extracting conformal data from the lattice}
\label{sec:extraction}

In this section we discuss how to extract conformal data by computing matrix elements of the operators $H_n$ of~\eqref{eq:Hn} between low-energy states $\ket{\varphi_\alpha}$. Here, each state $\ket{\varphi_\alpha}$ is a simultaneous eigenstate of the (normalized) critical lattice Hamiltonian $H$ and of the lattice momentum operator $P$ or, more precisely, of the lattice translation operator $e^{i\frac{2\pi}{N} P}$ that implements a translation by one lattice site,
\begin{equation} \label{eq:latticeHP}
 H \ket{\varphi_n} = E_\alpha \ket{\varphi_\alpha},~~~~~
 e^{i\frac{2\pi}{N} P}\ket{\varphi_\alpha}= e^{i\frac{2\pi}{N} S_\alpha}\ket{\varphi_\alpha}.
 \end{equation} 

We assume that, on these low-energy states, $H_n$ acts analogously to $H^\CFT_n$ of~\eqref{eq:Hn_L_CFT}, up to finite-size corrections that decrease with the size $N$ of the lattice. 

\subsection{Normalization of $H$ and central charge $c$}
\label{subsec:normalization}

So far we have assumed that the critical lattice Hamiltonian $H$ was already normalized so that its spectrum is given by \eqref{eq:FS_spec} (or, equivalently, so that the speed of light equals $1$ in the large-$N$ limit). However, in general the input data may be an unnormalized critical Hamiltonian $\tilde{H}$ or, equivalently, an unnormalized Hamiltonian density $\tilde{h}_j$, which relate to the normalized $H$ and $h_j$ through
\begin{equation} \label{eq:H_unnormalized}
H = a\tilde{H} + N b, ~~~~~~~h_j = a\tilde{h}_j + b,
\end{equation}
where $a$ and $b$ are two model-dependent constants. The constant $b$ can be computed by requiring that the extensive part of the ground state energy vanish in the limit of large $N$ (via a large-$N$ extrapolation), but in the following we will be able to simply ignore it, mostly because $b$ does not affect operators $H_n$ for $n\not = 0$. 

For a given system size $N$, the constant $a$ can be determined using states that are present in, and relations that are valid for, \emph{any} CFT (see Sect.~\ref{sec:ConformalTowers}). First we identify the states $\ket{\eye}$ and $\ket{T}$ as eigenstates of $\tilde{H}$
\begin{equation}
 \tilde{H} \ket{\eye} = \tilde{E}_\eye \ket{\eye},~~~~\tilde{H} \ket{T} = \tilde{E}_T \ket{T},
 \end{equation} 
 such that $\ket{\eye}$ is the unique ground state of $\tilde{H}$ and $\ket{T}$ is the eigenstate with momentum $P_T = 2 \times \frac{2\pi}{N} $ that has maximal overlap with $\tilde{H}_{-2} \ket{I}$ (where $\tilde{H}_{-2} $ is defined as $H_{-2}$ in \eqref{eq:Hn} after replacing $h_j$ with $\tilde{h}_j$). This last identification is motivated by the CFT relation \eqref{eq:TfromIH2}. Then, recalling that the scaling dimension of $T$ is $\Delta_T=2$, and therefore $E^\CFT_T -E^\CFT_\eye = \Delta_T \times \frac{2\pi}{N}  = 2 \times \frac{2\pi}{N}$, we set $a = \frac{4\pi}{N}/(\tilde{E}_T - \tilde{E}_\eye)$, since this guarantees that the (normalized) lattice energies also fulfill $E_T - E_\eye = 2 \times \frac{2\pi}{N}$.
 
With this normalization of $H$ the energies and momenta on the lattice read
\begin{equation}
\Delta_\alpha \approx \frac{N}{2\pi} \left(E_\alpha - E_\eye\right),  ~~~S_\alpha = \frac{N}{2\pi} P_\alpha,
\end{equation}
as we wanted. We can now estimate the scaling dimensions and conformal spins. Note: In the remainder (particularly Sect.~\ref{sec:results}), we slightly abuse notation, writing $H$ and $H_n$ for both the unnormalized and normalized operators. All results presented are obtained using the properly normalized versions.

Once we have normalized $h_j$, we can build the normalized Fourier modes $H_n$ using~\eqref{eq:Hn}. Through the relation~\eqref{eq:TfromIH2}, the central charge $c$ of the emergent CFT can then be estimated by the simple expectation value \cite{koo_representations_1994}
\begin{equation} \label{eq:c_lat}
c \approx 2\bra{\eye} H_2^\dagger H_2 \ket{\eye}.
\end{equation}
Alternatively, in order to eliminate finite-size corrections of $H_2$ that connect $|\eye\rangle$ to states other than $|T\rangle$, we can use
\begin{equation} \label{eq:c_lat2}
  c \approx 2 |\langle T | H_2 | \eye \rangle|^2,
\end{equation}
which often produces more accurate results. In either case, an extrapolation to large $N$ increases the accuracy of the lattice estimate of the central charge $c$.

The above procedures to normalize $H$ and estimate $c$ differ from previous proposals in that here we use $H_2$. The usual procedure to normalize $H$ is to identify $\ket{T}$ as the lowest-energy state with $P_\alpha = 2\times  \frac{2\pi}{N}$ ~\cite{\Henkel}. However, this fails if finite-size corrections shift the energy of another state with $P_\alpha = 2\times  \frac{2\pi}{N}$ below that of $|T\rangle$, as happens e.g.\ in the ANNNI model discussed in Sect.~\ref{sec:res_ANNNI}. Finally, an important advantage of estimating $c$ using $H_2$, compared to an extrapolation using the ground state energy alone \cite{\Henkel}, is that the latter also requires an extrapolation of the nonzero extensive contribution to the ground state energy, represented by $b$ in~\eqref{eq:H_unnormalized}, which must be subtracted before attempting to extrapolate $c$. 

\subsection{Primary states and conformal towers}

We now propose a criterion to identify \emph{candidates} for primary states. 
In the CFT, primary states obey~\eqref{eq:PstatesHn}. In words, they are the states that cannot be descended to lower energies by $H_n^{\CFT}$ or $\overline{H}{}^\CFT_n$. 
On the lattice at finite $N$ we have corrections to the energies \eqref{eq:FS_spec} and to the $H_n$, both of which must be allowed for in defining a criterion to identify candidates for a primary state. That is, on the lattice we need an approximate version of \eqref{eq:PstatesHn}.

To this end, we define $\epsilon^{(n)}$ to be the norm of the matrix elements of $\frac{1}{2}(H_{+n} + H_{-n})$ that connect an energy eigenstate $|\varphi\rangle$ with states of lower energy:
\begin{align} \label{eq:epsn} 
   \epsilon^{(n)}_\varphi \equiv \left| \Gamma_{\varphi} \; \frac{H_{n} + H_{-n}}{2}\; | \varphi \rangle \right|, \quad \text{for } n=1,2.
\end{align}
We then define a primary candidate as a state with small $\epsilon^{(1)}$ and $\epsilon^{(2)}$:
\begin{multline} \label{eq:Pstates_Hn}
  |\varphi\rangle \text{ primary candidate} \Leftrightarrow \epsilon^{(1)}_\varphi + \epsilon^{(2)}_\varphi \le \epsilon_{\max},
\end{multline}
which is analogous to \eqref{eq:PstatesHn} for $\epsilon_{\max} = 0$. 

Having identified primary candidate states, we can build their conformal towers by applying sequences of $H_n$ to them. By matching such sequences with sequences of $L^\CFT_n, \overline{L}{}^\CFT_n$, taking finite-size corrections into account, we can then identify each nonprimary lattice eigenstate with a particular descendant state of the CFT. 

However, if we only want to know which conformal tower each nonprimary state belongs to, it suffices to examine the matrix elements of a single operator -- one that connects each primary state with \emph{all} its descendants. We saw in Sect.~\ref{sec:ConformalTowers} that sequences of the ladder operators $L^\CFT_{-1}$, $L^\CFT_{-2}$, $\overline{L}{}^\CFT_{-1}$, and $\overline{L}{}^\CFT_{-2}$ acting on the primary are enough to reach any descendant in the CFT. On the lattice we can therefore use the matrix elements
\begin{align} \label{eq:intower_Hn}
  \tau_{\varphi'}^{\varphi} \equiv |\langle \varphi'| e^{\ic( H_1^{\diamond} + H_2^{\diamond} + H_{-1}^{\diamond} + H_{-2}^{\diamond} )} |\varphi\rangle|,
\end{align}
where $H_n^{\diamond}$ is the projection of $H_n$ onto the numerically obtained low-energy subspace and the exponential generates all sequences of $H_{\pm 1,\pm 2}$ (note that $H_n^\dagger = H_{-n}$). We then assign a nonprimary state $|\varphi'\rangle$ to the tower of the primary candidate $|\varphi\rangle$ that maximizes $\tau_{\varphi'}^{\varphi}$. Note that this procedure is suboptimal in the sense that finite-size corrections accumulate when we take products of $H^{\diamond}_n$. More sophisticated schemes avoiding this issue are possible \cite{zou_conformal_2017}, but this simpler scheme is already sufficient for our purpose of illustrating the usefulness of $H_n$.

Armed with an identification of each eigenstate of $H$ at fixed $N$, we may examine data from a range of sizes to determine if the assignment is robust. To check that the identification of primary states is robust we note that, using~\eqref{eq:Pstates_Hn}, we can verify statements such as ``With $\epsilon_{\max} = 10^{-6}$ there is a primary candidate at $\Delta \approx 3$ and $S=3$ for all tested system sizes $N \ge 6$''. Since finite-size corrections typically obey power-law or logarithmic scaling in the system size \cite{cardy_operator_1986, cardy_logarithmic_1986}, we rely on them varying smoothly with $N$ and assume that primary candidate states $|\varphi\rangle_N$ at different $N$, but with similar energy and the same momentum, represent the same primary operator in the CFT. For such sequences of primary candidate states we should find that both $\epsilon^{(1)}_\varphi(N)$ and $\epsilon^{(2)}_\varphi(N)$ go to zero in the limit of large~$N$.

\subsection{Quasiprimaries and global conformal towers }

The identification of primary states on the lattice, as discussed above, is a central application of the correspondence between the CFT Fourier modes $H^{\CFT}_n$ and their lattice analogues $H_n$ \cite{koo_representations_1994}, because of its direct impact on our ability to compute the conformal data of the underlying CFT, which requires such an identification. However, a more refined characterization within each conformal tower is also possible on the lattice, as we discuss next.

A conformal tower (or \emph{Virasoro} tower) decomposes into infinitely many \emph{global} conformal towers, each consisting of a quasiprimary operator and its global descendants. To identify quasiprimary states on the lattice, we resort to an approximate version of \eqref{eq:QPstatesHn} in terms of the error $\epsilon^{(1)}_\varphi$ defined in \eqref{eq:epsn}, namely
\begin{multline} \label{eq:QPstates_Hn}
  |\varphi\rangle \text{ quasiprimary candidate} \Leftrightarrow \epsilon^{(1)}_\varphi \le \epsilon_{\max},
\end{multline}
which indeed is analogous to \eqref{eq:QPstatesHn} for $\epsilon_{\max} = 0$. Then, once a quasiprimary state $\ket{\varphi}$ has been identified, its global conformal tower (generated in the CFT by acting on $\ket{\varphi}$ with powers of $L^{\CFT}_{-1}$ and $\overline{L}{}^{\CFT}_{-1}$ or, equivalently, powers of $H_1^{\CFT}$ and $H_{-1}^\CFT$) can be produced by studying the matrix elements
\begin{align} \label{eq:inGlobal_Hn}
  \kappa_{\varphi'}^{\varphi} \equiv |\langle \varphi'| e^{\ic (H_1^{\diamond} + H_{-1}^{\diamond}) } |\varphi\rangle|,
\end{align}
where $H_1^{\diamond}, H_{-1}^{\diamond}$ are defined above and similar considerations to \eqref{eq:intower_Hn} apply. 

\section{Results}
\label{sec:results}

\subsection{The Ising model}
\label{sec:res_ising}

\begin{figure}
\includegraphics[width=\linewidth]{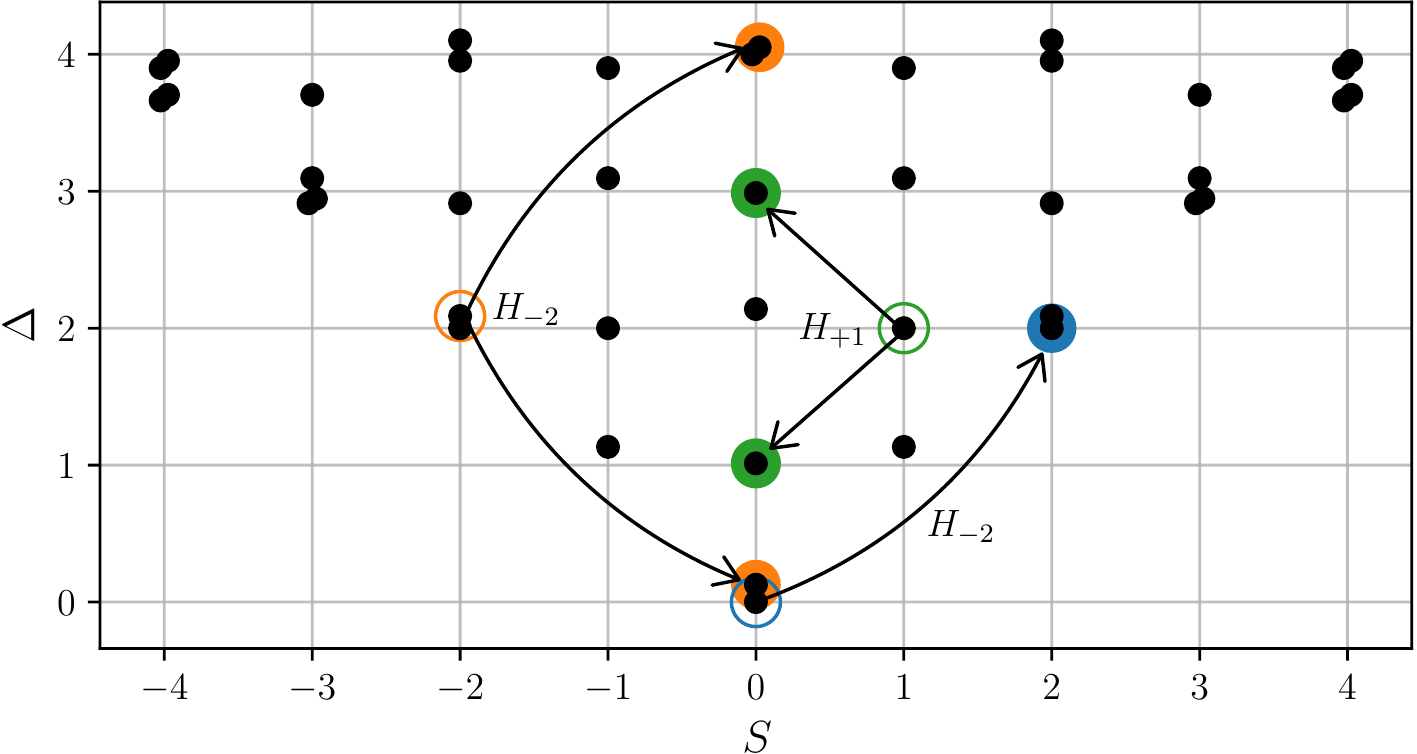}
\caption{\label{fig:ising_H_action} Spectrum of the Ising model at system size $N=14$ with energies and momenta in terms of $\Delta$ and $S$, showing the action of $H^\TFI_{+1}$ and $H^\TFI_{-2}$ on selected energy eigenstates. The empty circles identify the states $|\varphi_\alpha\rangle$ to which the operator is applied and the filled circles indicate the sizes of the matrix elements $\langle \varphi_\beta | H^\TFI_n | \varphi_\alpha \rangle$  with the remaining eigenstates $|\varphi_\beta\rangle$, on a logarithmic scale. Very small matrix elements $< 10^{-12}$ are not plotted.
}
\end{figure}

\begin{figure}
\includegraphics[width=\linewidth]{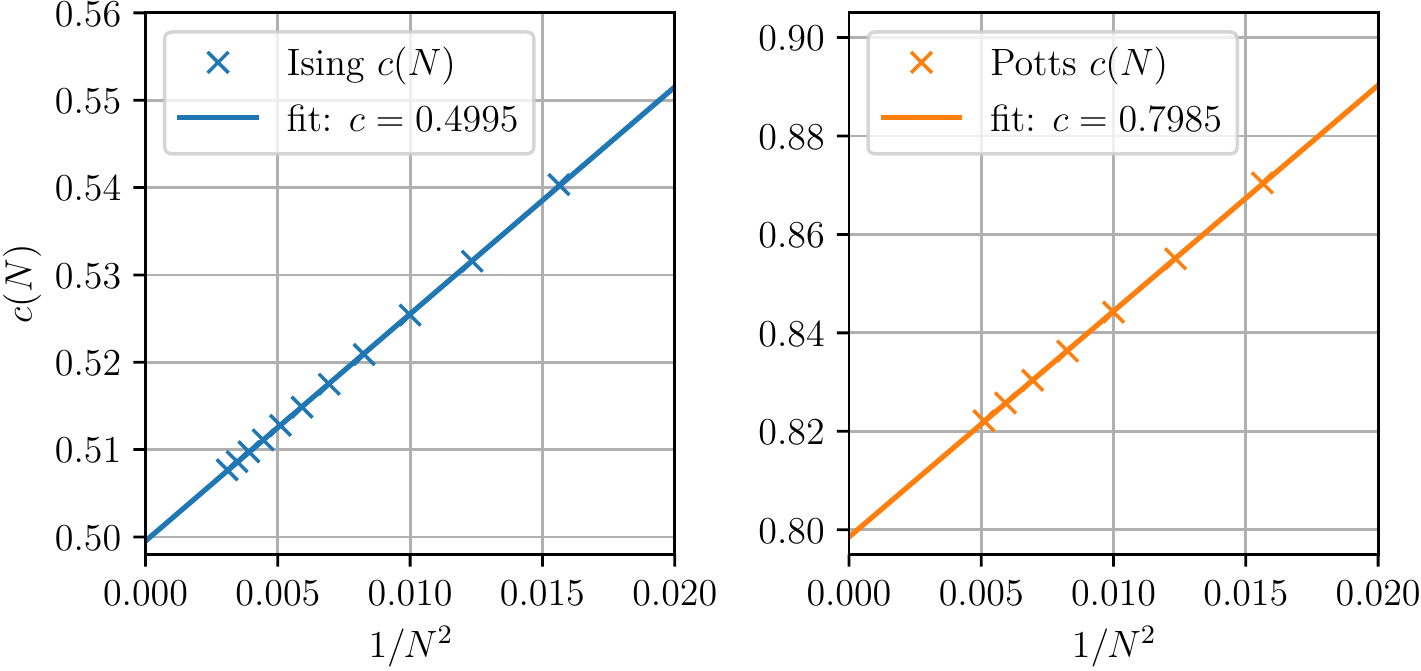}
\caption{\label{fig:c_scaling} Central charge from~\eqref{eq:c_lat}, with linear extrapolation to large~$N$ using all visible data. System sizes shown are $N=8\dots 18$ for the Ising model and $N = 8 \dots 14$ for the three-state Potts model. We do not provide an error for the extrapolated~$c$ since there are systematic finite-size corrections on each point. The scaling exponent~$2$ is consistent with known finite-size corrections present in both models~\cite{cardy_operator_1986, henkel_finite-size_1987, reinicke_analytical_1987}.
}
\end{figure}

As a first test of the methods introduced in Sect.~\ref{sec:extraction}, we examine the behavior of the Hamiltonian density modes $H_n$ for the integrable transverse field Ising model of~\eqref{eq:H_TFI}, for which some conformal data was extracted in \cite{koo_representations_1994}. The Hamiltonian is invariant under a global spin flip $\prod_{j=1}^N \sigma_j^Z$, and is critical at its self-dual point $\lambda = 1$ \cite{\Henkel}.

We construct $H_n$ for the critical Ising model as
\begin{align} \label{eq:Hn_Ising}
  H^\TFI_n \equiv -\frac{N}{2\pi} \sum_{j=1}^N \left( e^{\ic j n \frac{2\pi}{N}} \sigma^Z_j + e^{\ic (j+\frac{1}{2}) n \frac{2\pi}{N}} \sigma^X_j \sigma^X_{j+1}\right),~~~
\end{align}
where we have chosen different phases for the onsite terms $\sigma^Z$ and the bond terms $\sigma^X\sigma^X$ to reflect that the bonds are centered \emph{between} two sites. We propose \emph{in general} that terms with support on sites $j$ and $j+r$, and optionally the sites in between, be given phases consistent with the midpoint $x=j+r/2$. For the Ising model, this ensures that $H^\TFI_n$ remains consistent with Kramers-Wannier duality, which exchanges the $\sigma^X_j \sigma^X_{j+1}$ and $\sigma^Z_j$ terms.

For a given finite system size $N$, we simultaneously diagonalize the Hamiltonian and the translation operator, with periodic boundary conditions, using the Arnoldi algorithm -- a Krylov-subspace method for finding eigenvalue/eigenvector pairs of nonhermitian matrices \cite{arnoldi_principle_1951} -- to find a set of low-energy eigenstates $|\varphi_\alpha\rangle$, with energies $E_\alpha$ and momenta $P_\alpha$. In this case, we compute the 41 lowest-energy eigenvalues and corresponding eigenvectors. With these we compute the matrix-elements $\langle \varphi_\beta | H_n^\TFI | \varphi_\alpha \rangle$ in the low-energy eigenbasis of $H$, which we normalize according to the discussion in Sect.~\ref{subsec:normalization}.

For our first test of the behavior of $H^\TFI_n$, we examine a selection of matrix elements for $n = \pm 1,2,3$. We find that the action of these $H^\TFI_{n}$ within the computed basis of 41 low-energy states is indeed consistent with that of their CFT counterparts~\eqref{eq:Hn_L_CFT}, described in Sect.~\ref{sec:ConformalTowers}, as expected from~\cite{koo_representations_1994}. In particular, despite noticeable finite-size corrections to the energies, states $H^\TFI_n|\varphi_\alpha\rangle$ have nonzero overlap \emph{only} with energy eigenstates of scaling dimension $\Delta_\alpha \pm n + \mathcal{O}(\epsilon)$ (where $\epsilon \ll 1$ represents finite-size corrections to the energies) and spin $S_\alpha - n$, as expected from the CFT result of~\eqref{eq:HnCFT_action}. Overlaps with states of incompatible scaling dimension are zero \emph{to numerical precision} (within the 41 low-energy states under consideration). We plot a few examples in Fig.~\ref{fig:ising_H_action}.

\begin{figure}
\includegraphics[width=\linewidth]{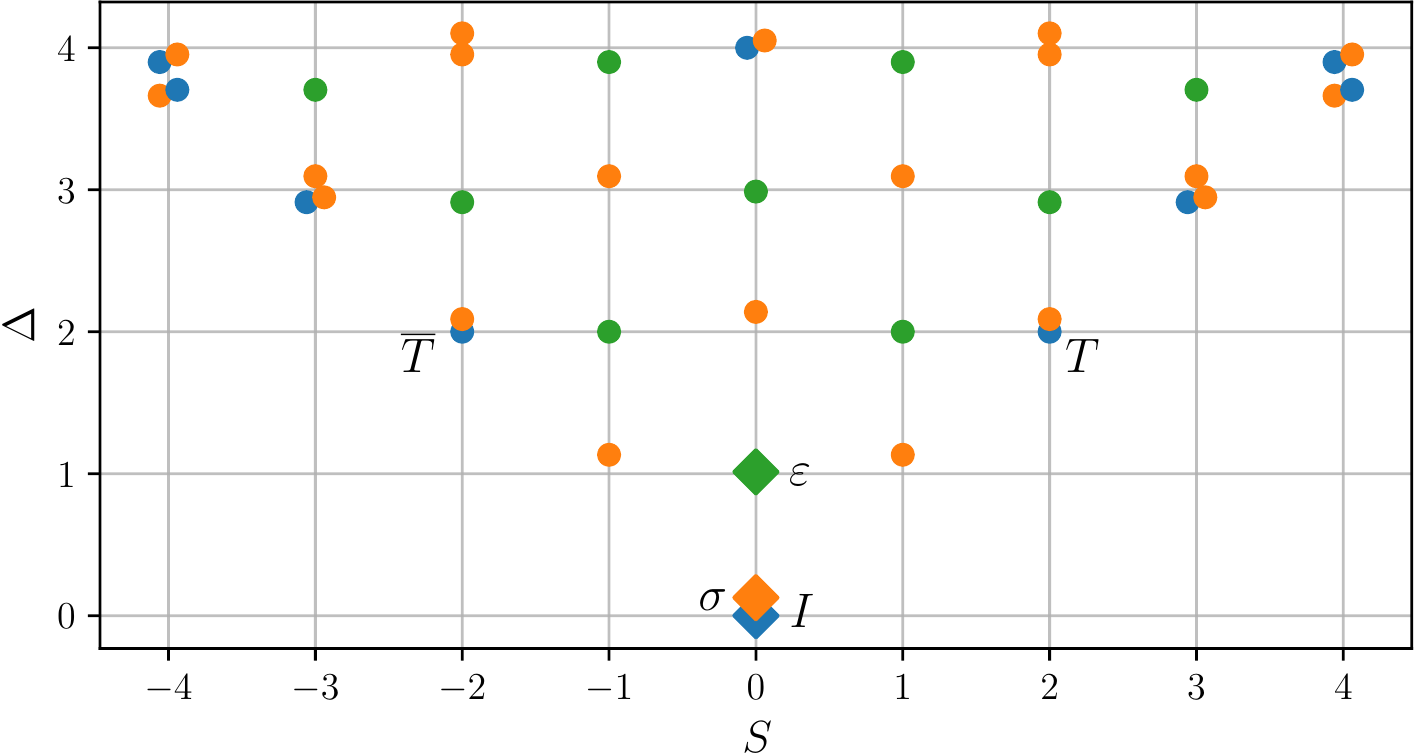}
\caption{\label{fig:ising_lattice_spec} Ising model spectrum at system size \mbox{$N=14$}, with energies and momenta in terms of $\Delta$ and $S$. States are colored according to their numerically identified conformal towers. Primary candidate states, identified using~\eqref{eq:Pstates_Hn} with $\epsilon_{\max} = 10^{-14}$, are marked with diamonds. 
}
\end{figure}

\begin{figure}
\includegraphics[width=\linewidth]{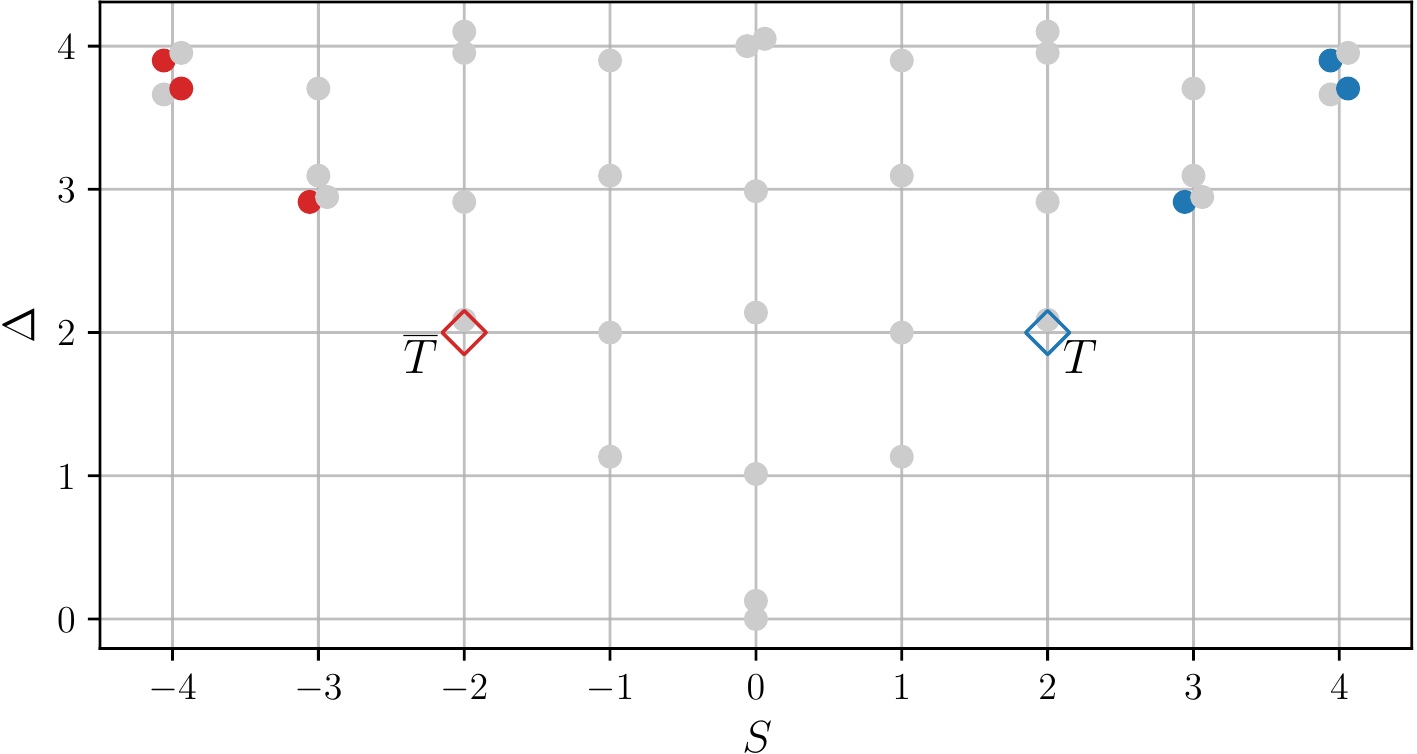}
\caption{\label{fig:ising_global_towers} Ising model spectrum at system size \mbox{$N=14$} showing two quasiprimary states $|T\rangle$ and $|\overline{T}\rangle$ (empty diamonds) determined from~\eqref{eq:QPstates_Hn}. The colored dots are states connected to each quasiprimary according to~\eqref{eq:inGlobal_Hn}. Most of these correspond to global descendants of the CFT operators $T$ and $\overline{T}$. However, there is a \emph{linear combination} of the two blue (red) states with $S=4$ ($S=-4$) that fulfills~\eqref{eq:QPstates_Hn} and thus corresponds to a \emph{quasiprimary} CFT operator. See App.~\ref{app:degenerate_qp}.
}
\end{figure}

Applying~\eqref{eq:Pstates_Hn} to determine the primary candidate states, we find that, even at $N=14$, we can correctly identify all three primary states using a tolerance close to machine precision, $\epsilon_{\max} = 10^{-14}$. Although it is trivial that the primary states in the Ising model cannot be lowered in energy (there are no states at compatible momenta that they could be lowered to), it is nontrivial, if unsurprising, that no descendant states (again, within the 41 low-energy states under consideration) are \emph{misidentified} as primary. That said, later we will see that the Potts model provides a much better proving ground for the identification of primary states.

We further observe that $\tau_{\varphi'}^\varphi$ of~\eqref{eq:intower_Hn} delivers a completely unambiguous tower assignment to the remaining states, consistent with the observed perfect ladder behavior of $H^\TFI_n$. In other words, there are no significant finite-size corrections that mix conformal towers. Indeed, in this case such corrections are disallowed by the symmetries of $H_n^\TFI$ (this is not the case for the Potts model -- see below).

Corrections are present, however, which affect the size of the \emph{nonzero} matrix elements of $H^\TFI_n$, as evidenced by the central charge estimates obtained from~\eqref{eq:c_lat} shown in Fig.~\ref{fig:c_scaling}. Nevertheless, we obtain excellent agreement with $c=\frac{1}{2}$ after extrapolation to large~$N$, in concordance with the results of~\cite{koo_representations_1994}.

Fig.~\ref{fig:ising_lattice_spec} shows the identification of eigenstates with primary operators and their descendants at system size $N=14$. Comparing with the Ising CFT spectrum of Fig.~\ref{fig:ising_towers} we observe that, even in cases of very significant finite-size corrections to the energies, preventing an identification of the tower using the spectrum alone, we are able to use $H^\TFI_n$ to make an unambiguous identification.

The identification of global conformal towers using $\kappa_{\varphi'}^\varphi$ of~\eqref{eq:inGlobal_Hn} was equally successful, as demonstrated in Fig.~\ref{fig:ising_global_towers}.

\subsection{Three-state Potts model}
\label{sec:potts}

We now test our algorithms with the three-state Potts model, which has a more complicated emergent CFT hosting more primary operators than the Ising CFT, including ones with significantly larger scaling dimensions. These are much harder to characterize numerically, partly because finite-size corrections to the $H_n$ operators mix conformal towers, as detailed below.  

The three-state Potts model \cite{wu_1982} may be thought of as a generalization of the Ising model in which spins have not two positions (up and down),  but three. Unlike the Ising model it is not equivalent to a theory of free particles. It is, however, integrable at criticality \cite{fateev_self-dual_1982}. The Hamiltonian
\begin{align} \label{eq:Hpotts}
  H^{\potts} (\lambda) \equiv -\frac{1}{2}\sum_{j=1}^N \left[U_j U_{j+1}^\dagger + \lambda V_j \right] + \text{h.c.}
\end{align}
has a critical point at $\lambda = 1$, determined by self-duality, and may be represented in terms of matrices
\begin{align}
{U}=\begin{pmatrix}
1& 0& 0 \\
0& \omega & 0 \\
0& 0& \omega^*
\end{pmatrix}, \quad
{V}=
\begin{pmatrix}
0& 0& 1\\
1& 0& 0\\
0& 1& 0
\end{pmatrix}, \quad
{\omega} = e^{\ic \frac{2\pi}{3}},
\end{align}
which obey the exchange relations
\begin{align}
  UV = \omega VU.
\end{align}
The Hamiltonian is manifestly invariant under the global shift $\prod_{j=1}^N V_j$, which implies that eigenstates fall into one of three $\mathbb{Z}_3$ charge sectors. At criticality its low-energy physics is described by the three-state Potts CFT, which has $c=4/5$ and twelve primary operators, including some with nonzero spin and four with scaling dimension $\Delta > 2$ \cite{cardy_operator_1986, gehlen_operator_1986}, making their identification nontrivial. The eight primary operators of the $\mathbb{Z}_3$ \emph{zero-charge} sector are:
\begin{center}
  \renewcommand{\arraystretch}{1.2}
  \setlength{\tabcolsep}{5pt}
\begin{tabular}{r|cccccccc}
  \toprule[0.8pt]
       & $\eye$ & $\varepsilon$ & $\Phi_{\varepsilon\overline{X}}$ & $\Phi_{X\overline{\varepsilon}}$ & $X$    & $\overline{W}$ & $W$ & $Y$ \\ \midrule[0.4pt]
  $\;\Delta\;$ & $0$    & $4/5$         & $9/5$                         & $9/5$                         & $14/5$ & $3$            & $3$ & $6$ \\ 
  $\;S\;$      & $0$    & $0$           & $-1$                          & $+1$                          & $0$    & $-3$           & $+3$ & $0$ \\
  \bottomrule[0.8pt]
\end{tabular}
\end{center}
Here, we have largely followed the notation of \cite{mong_parafermionic_2014}.

\begin{figure}
\includegraphics[width=\linewidth]{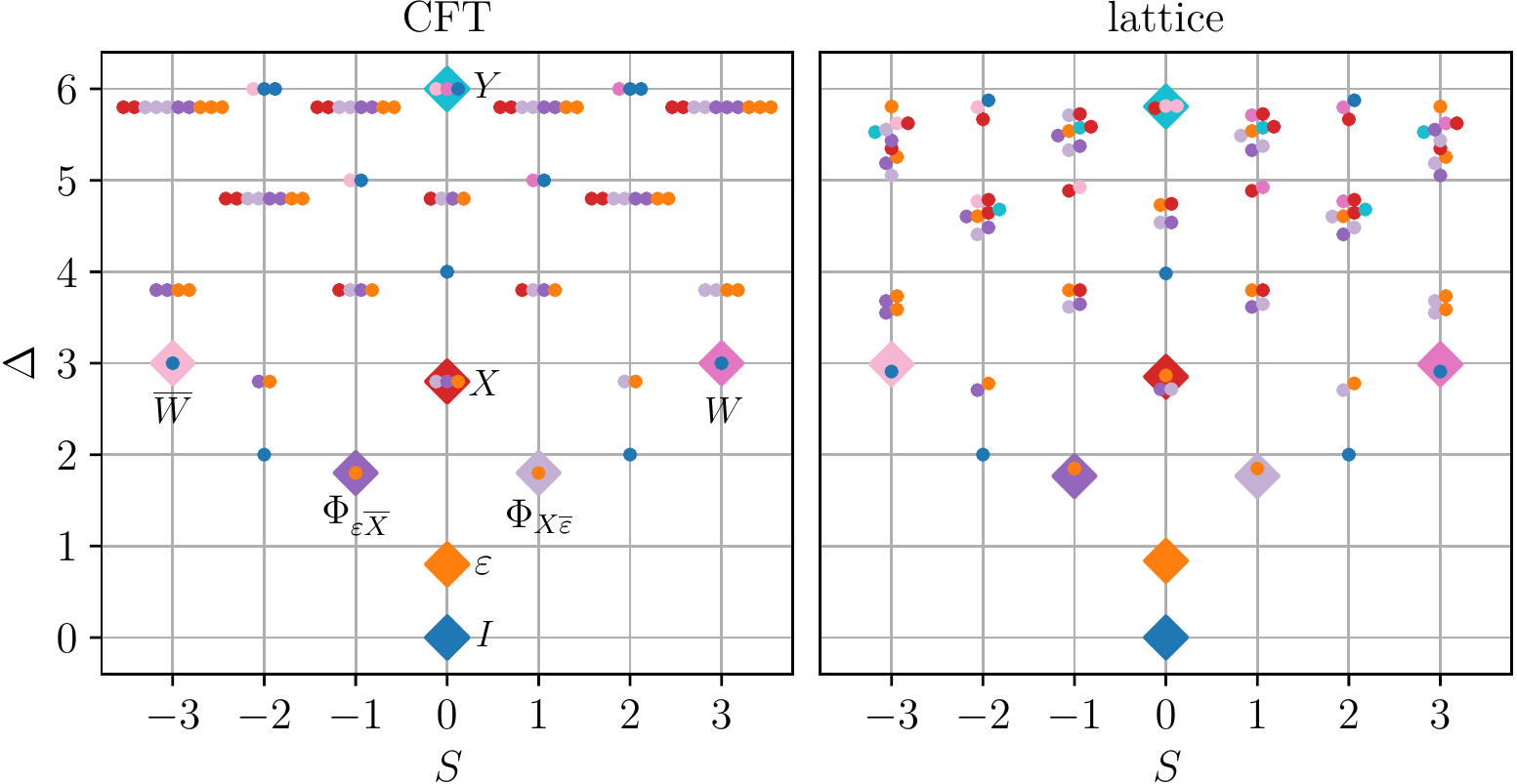}
\caption{\label{fig:potts_spec} Three-state Potts CFT spectrum with labeling of the primaries (left) and lattice spectrum at system size \mbox{$N=14$} (right). We restrict to the zero $\mathbb{Z}_3$ charge sector. Lattice primaries and descendants are identified as in Fig.~\ref{fig:ising_lattice_spec} using a tolerance $\epsilon_{\max} = 0.2$ for primaries. For $\Delta>3$ we restrict to spins \mbox{$|S| \le 3$}, allowing numerical identification of primaries with $|S| \le 1$.
We see that even high-$\Delta$ and chiral ($S \neq 0$) primaries are identified successfully in the lattice data, and that towers are mostly consistent with the CFT, despite the simplicity of the algorithm used for tower identification (see Sect.~\ref{sec:extraction}). See main text for a discussion of errors.
}
\end{figure}

We first define the Hamiltonian density modes
\begin{multline} \label{eq:Hn_Potts}
  H^\potts_n \equiv -\frac{N}{2\pi} \sum_{j=1}^N \left[ e^{\ic j n \frac{2\pi}{N}} (V_j + \text{h.c.}) + \right. \\
  \left. e^{\ic (j+\frac{1}{2}) n \frac{2\pi}{N}} (U_j U^\dagger_{j+1} + \text{h.c.}) \right],
\end{multline}
using them with the algorithms of Sect.~\ref{sec:extraction} to determine primary candidates and tower assignments.

\begin{figure}
\includegraphics[width=1.0\linewidth]{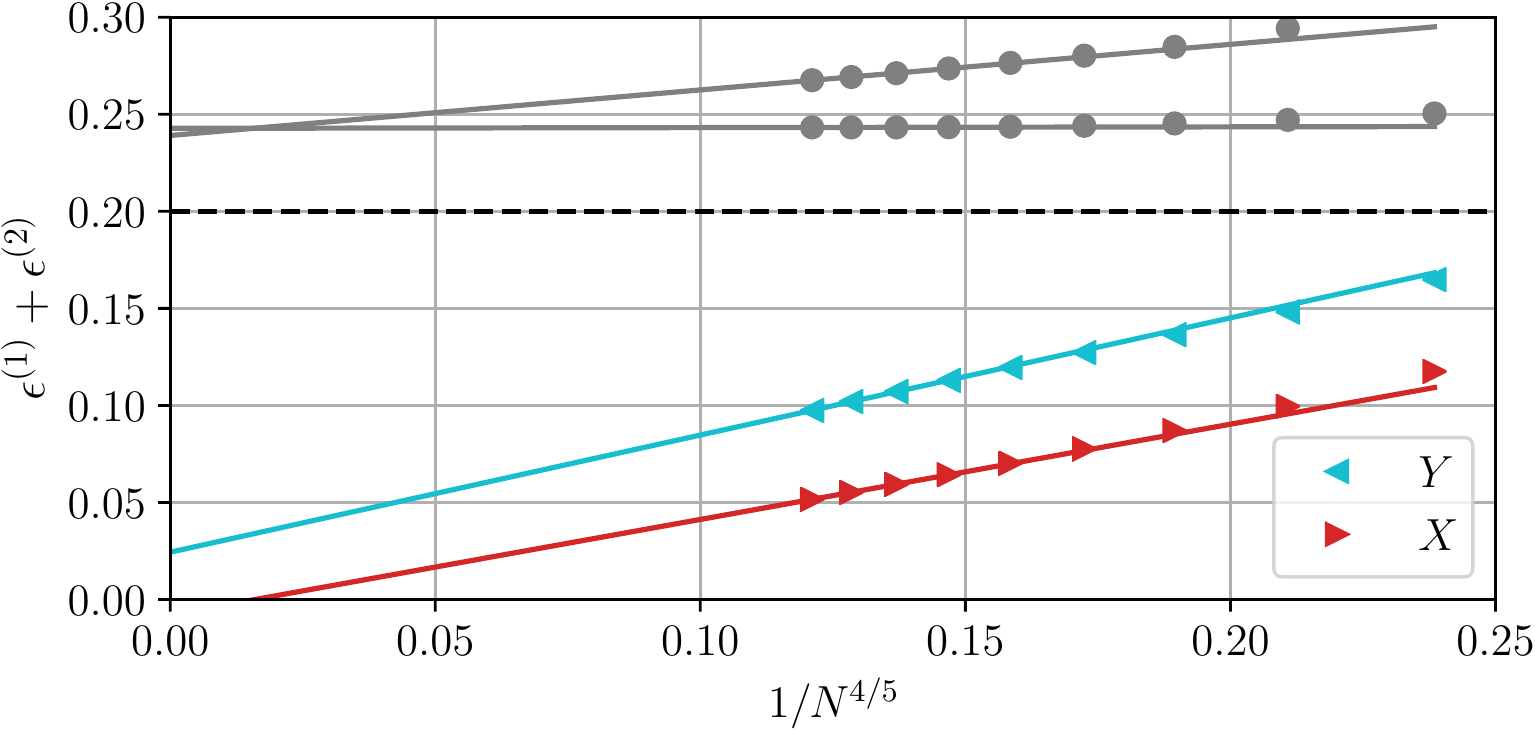}
\caption{\label{fig:potts_pri_err} Scaling with system size $N$ of matrix elements of $H^\potts_1$ and $H^\potts_2$ that lower the energy of the $X$ and $Y$ primary candidate states, quantified using~\eqref{eq:epsn}. The dashed line marks the threshold $\epsilon_{\max} = 0.2$ used to distinguish primaries from descendants in Fig.~\ref{fig:potts_spec}. Using linear regression on the four leftmost points, we see these matrix elements appear to vanish in the large-$N$ limit, consistent with these being primary states. For comparison, we show the scaling for two descendant states in gray. The scaling exponent $4/5$ is consistent with the leading finite-size correction in the Potts model~\cite{gehlen_conformal_1987, reinicke_analytical_1987}.
}
\end{figure}

At system size $N=14$ we are able to use~\eqref{eq:Pstates_Hn} to identify \emph{all eight} primary states of the charge-zero sector, as shown in Fig.~\ref{fig:potts_spec}, albeit at a relatively high tolerance $\epsilon_{\max} = 0.2$. This is needed because, although we find $\epsilon^{(1)}$ to be negligible for all primary candidate states (marking them unambiguously as \emph{quasiprimary} states), $\epsilon^{(2)}$ is significant for the $X$ and $Y$ primary candidates due to matrix elements of $H^\potts_2$ connecting those states to lower-energy states. To justify setting $\epsilon_{\max} = 0.2$ to suppress these matrix elements, we must examine their scaling with $N$. In Fig.~\ref{fig:potts_pri_err} we show that $\epsilon^{(2)}_X(N)$ and $\epsilon^{(2)}_Y(N)$ both appear to go to zero in the large $N$ limit, confirming the assignment of these lattice states to the $X$ and $Y$ primary operators. The scaling exponent $4/5$ used in Fig.~\ref{fig:potts_pri_err} is that of the known leading finite-size correction of the Potts model \cite{gehlen_conformal_1987, reinicke_analytical_1987}. 

We note that identification of primaries is generally \emph{not possible} using only the spectral data since there may be lower-energy states which, from their energies and momenta at finite size alone, cannot be excluded from being in the same tower as the primary state. That we can confidently identify all primaries in the Potts model, including at large~$\Delta$, thus demonstrates a key benefit of using $H_n$ to extract conformal data.

Finite-size corrections to $H^\potts_n$ at $N=14$ also affect identification of conformal towers using~\eqref{eq:intower_Hn}. Comparing with the CFT spectrum in Fig.~\ref{fig:potts_spec} we find that, although most assignments are plausible, some of the higher-energy states are clearly misidentified. For example, the erroneous matrix elements of $H^\potts_2$ affecting the $Y$ primary lead to the misidentification of $\varepsilon$ descendants as belonging to the $Y$ tower. Furthermore, we find that elements of the identity tower are sometimes misidentified as $X$ descendants. Although the former could easily be eliminated if, when assigning towers to descendants, we only considered primaries with lower energies than the descendant, the latter could not. For more precision, tower assignment should be based on a finite-size scaling analysis similar to that of Fig.~\ref{fig:potts_pri_err}.

The tower-mixing errors we observe here are consistent 
with the known finite-size corrections to the eigenstate energies \eqref{eq:FS_spec} of the Potts model. These can be understood as coming from perturbations of the uncorrected CFT Hamiltonian density $h^\CFT(x)$ by \emph{irrelevant operators} (those with $\Delta > 2$) \cite{cardy_operator_1986}. Of course, such perturbations must also affect the Hamiltonian density Fourier modes $H_n$ and we can understand the nature of these corrections in terms of the \emph{operator algebra} \cite{\CFTped} of the CFT. In this case, perturbation of $h^\CFT(x)$ by the primary field operator $X(x)$ \cite{gehlen_conformal_1987, reinicke_analytical_1987} explains the mixing of the $X$ and $Y$ towers with the $\eye$ and $\varepsilon$ towers, respectively, in terms of the fusion rules $X \times X = \eye + X$ and $X \times Y = \epsilon$ of the Potts CFT operator algebra \cite{dotsenko_critical_1984}. As an aside for the interested reader, we also remark that the observed mixing connects different representations of the $\mathcal{W}_3$ algebra \cite{fateev_conformal_1987}, a symmetry of the three-state Potts CFT which includes the Virasoro algebra.

Finally, as for the Ising model, we obtain an accurate estimate of the central charge as shown in Fig.~\ref{fig:c_scaling}.

\subsection{The self-dual ANNNI model}
\label{sec:res_ANNNI}

\begin{figure}
  \includegraphics[width=\linewidth]{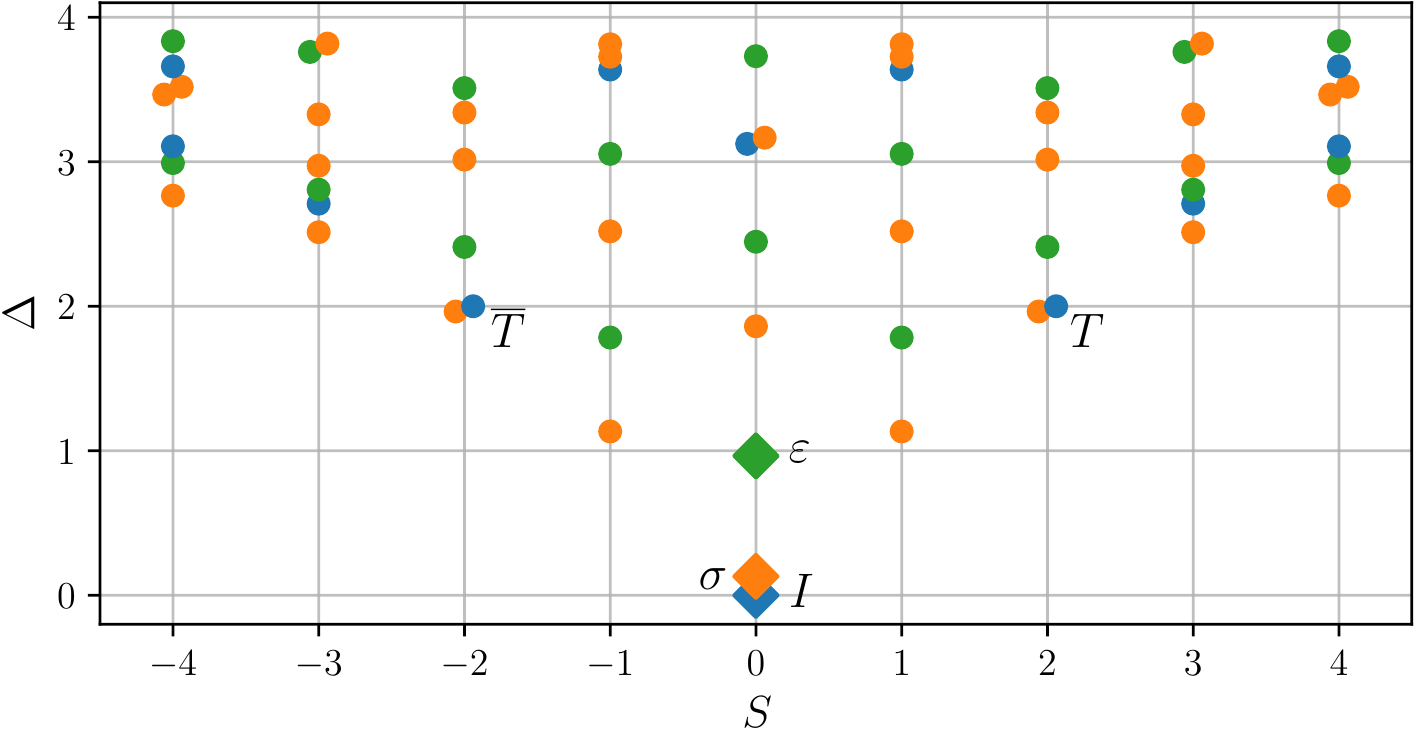}
\caption{\label{fig:ANNNI_pri} ANNNI model spectrum at $\gamma=0.5$ (nonintegrable) and system size $N=14$, with numerical identification of primary states and assignment of remaining states to conformal towers. Note that finite-size corrections to the energy are severe compared to Fig.~\ref{fig:ising_lattice_spec}, being sufficient to shift descendant states of $\sigma$ below the energy-momentum states $|T\rangle$ and $|\overline{T}\rangle$.
}
\end{figure}

\begin{figure}
\includegraphics[width=\linewidth]{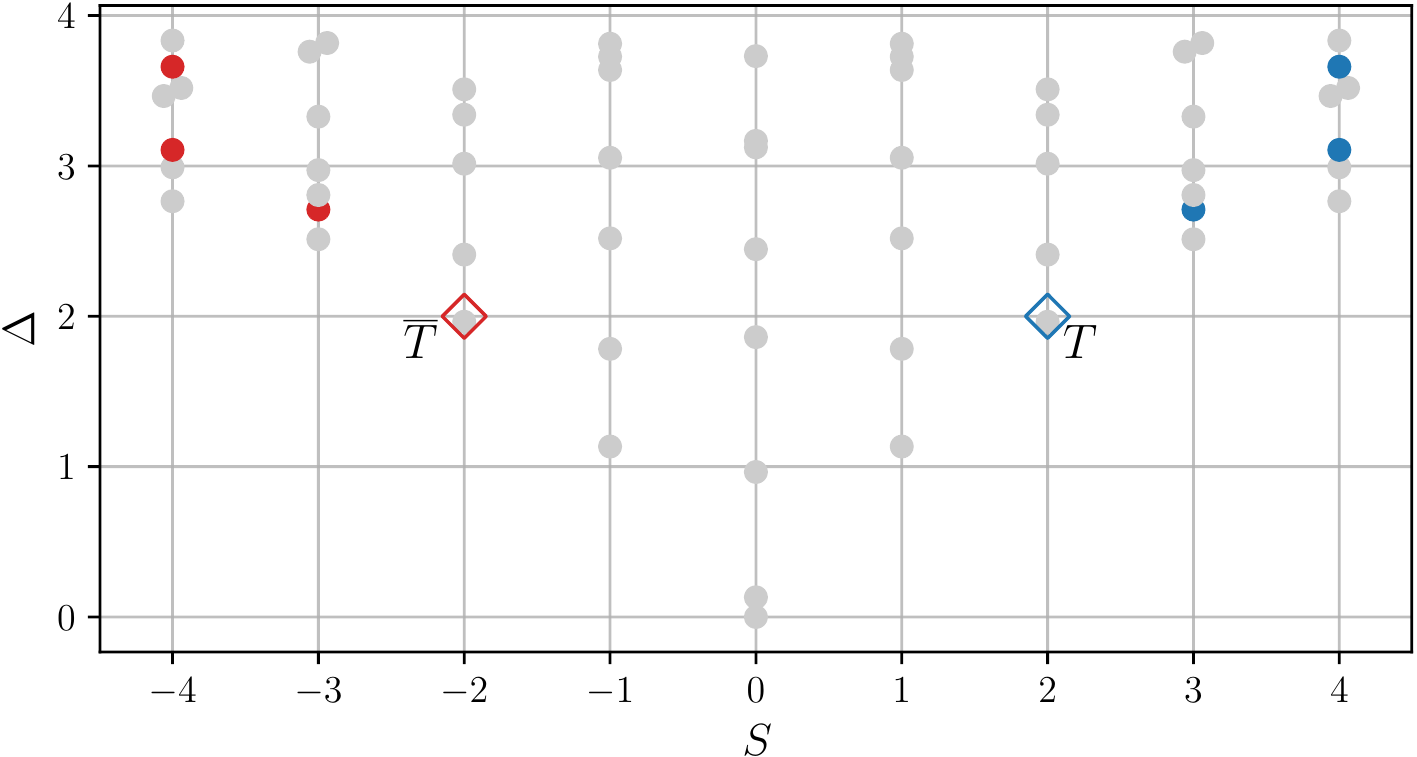}
\caption{\label{fig:ANNNI_global_towers} ANNNI model spectrum at $\gamma=0.5$ (nonintegrable) and system size \mbox{$N=14$} showing two quasiprimary states $|T\rangle$ and $|\overline{T}\rangle$ (colored empty diamonds) determined from~\eqref{eq:QPstates_Hn}. The colored dots are states connected to each quasiprimary according to~\eqref{eq:inGlobal_Hn}. Most of these correspond to global descendants of the CFT operators $T$ and $\overline{T}$. However, as for the Ising model, there is a \emph{linear combination} of the two blue (red) states with $S=4$ ($S=-4$) that fulfills~\eqref{eq:QPstates_Hn} and thus corresponds to a \emph{quasiprimary} CFT operator. See App.~\ref{app:degenerate_qp}.
}
\end{figure}

We are now ready to test the Koo-Saleur formula, as well as our conformal data extraction procedures using the Hamiltonian density Fourier modes $H_n$, for a nonintegrable system.
We consider the Axial Next-Nearest-Neighbor Ising (ANNNI) model \cite{selke_1988, milsted_statistical_2015, rahmani_phase_2015}, an extension of the Ising model~\eqref{eq:H_TFI} by a next-nearest-neighbor interaction term and its counterpart under duality, resulting in the Hamiltonian
\begin{align} \label{eq:H_ANNNI}
  H^\ANNNI = -\sum_{j=1}^N \left[ \sigma^X_j \sigma^X_{j+1} + \sigma^Z_j + \gamma \sigma^X_j \sigma^X_{j+2} + \gamma \sigma^Z_j \sigma^Z_{j+1} \right],
\end{align}
which with this parameterization is self-dual for all $\gamma$. Under a Jordan-Wigner transformation it becomes a translation-invariant chain of interacting Majorana fermion modes and in this context its phase diagram has recently been numerically examined \cite{milsted_statistical_2015, rahmani_phase_2015}. It was found to have two distinct gapless regimes within the (approximate) parameter range $-5 < \gamma < 250$, with an emergent Ising CFT for $-0.3 < \gamma < 250$.
We choose $\gamma = 0.5$, which is far from the critical Ising integrable point, but in a regime where the universality class is well understood, making the results easier to analyze.
We first compute the 71 lowest-energy eigenvectors of $H^\ANNNI(\gamma = 0.5)$, before evaluating the matrix elements in the low-energy eigenbasis of the Hamiltonian density Fourier modes, which we construct as
\begin{multline} \label{eq:Hn_ANNNI}
  H^\ANNNI_n \equiv -\frac{N}{2\pi} \sum_{j=1}^N \left[ e^{\ic j n \frac{2\pi}{N}} \left( \sigma^Z_j + \gamma \sigma^X_{j-1} \sigma^X_{j+1}\right) \right. \\
  \left. + e^{\ic (j+\frac{1}{2}) n \frac{2\pi}{N}} \left( \sigma^X_j \sigma^X_{j+1} + \gamma \sigma^Z_j \sigma^Z_{j+1} \right) \right],
\end{multline}
in the same way as we did for the Ising model in~\eqref{eq:Hn_Ising}.

\begin{figure}
\includegraphics[width=\linewidth]{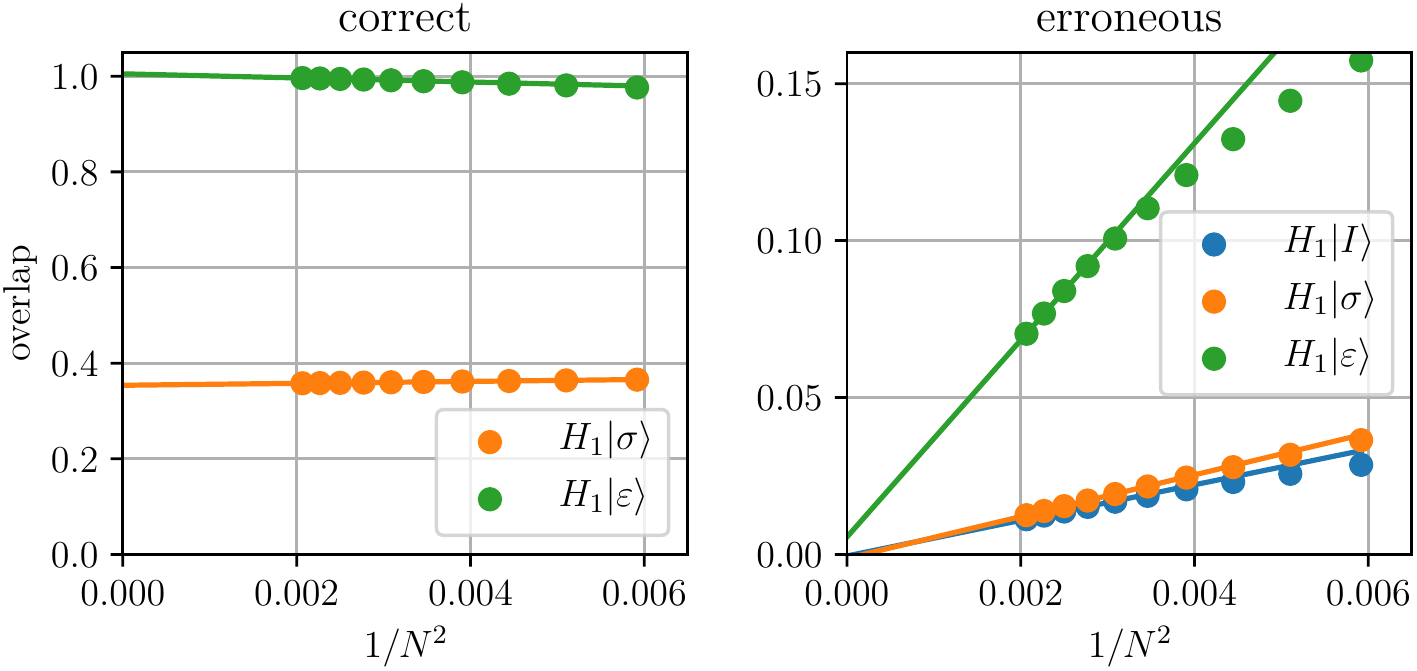}
\caption{\label{fig:ANNNI_err_overlaps} Scaling with system size $N$ of the overlaps of $H_1^\ANNNI |\eye\rangle$, $H_1^\ANNNI |\sigma\rangle$ and $H_1^\ANNNI |\varepsilon\rangle$ with correct (left) and erroneous (right) descendant states (from the same conformal tower as the primary). Data for $N=13 \dots 22$ is plotted. Linear regression is performed on the leftmost three points. No ``correct'' overlap is plotted for $H_1^\ANNNI |\eye\rangle$ since $H_1^\CFT |\eye\rangle=0$. We conclude that the erroneous overlaps are finite-size corrections that go to zero asymptotically as $1/N^2$.
}
\end{figure}

Although the model is not integrable, we obtain similar results to those of Sec.~\ref{sec:res_ising}. In particular we find that~\eqref{eq:Pstates_Hn} and~\eqref{eq:intower_Hn} deliver completely unambiguous identifications of primary states and conformal towers, which we plot in Fig.~\ref{fig:ANNNI_pri}. This is despite strong finite-size corrections to the energy eigenvalues compared to the Ising case of Fig.~\ref{fig:ising_lattice_spec}. We are also able to identify quasiprimary states and global descendants using~\eqref{eq:QPstates_Hn} and~\eqref{eq:inGlobal_Hn}, as we show in Fig.~\ref{fig:ANNNI_global_towers}.

However, corrections show up in the matrix elements of $H^\ANNNI_1$ and $H^\ANNNI_2$ that were not present in $H^\TFI_1$ and $H^\TFI_2$, for example we observe that $H^\ANNNI_1 |\eye\rangle$ has overlap with a state corresponding to a descendant of the $\eye$ operator with $\Delta = 5$, despite the CFT result $H^\CFT_1|\eye\rangle = 0$. Similarly, $H^\ANNNI_1 |\sigma\rangle$ has overlap with a state corresponding to a descendant of $\sigma$ with $\Delta = 3 \frac{1}{8}$, despite only one state with $\Delta = 1 \frac{1}{8}$ occurring as an overlap of $H^\CFT_1 |\sigma\rangle$ in the CFT, and $H^\ANNNI_1|\varepsilon\rangle$ has overlap with a state corresponding to an $\varepsilon$-descendant with $\Delta = 4$, in addition to the expected $\Delta = 2$. In order to justify calling these overlaps finite-size corrections, we must of course demonstrate that they disappear as $N\rightarrow \infty$. Using the examples from the $\sigma$ and $\varepsilon$ conformal towers mentioned above, we show in Fig.~\ref{fig:ANNNI_err_overlaps} that this is indeed the case.
We note that, as with the Ising model, there is no mixing of different conformal towers (again due to the symmetries of $H_n^\ANNNI$), explaining why we are still able to make tower assignments unambiguously.

\begin{figure}
\includegraphics[width=0.9\linewidth]{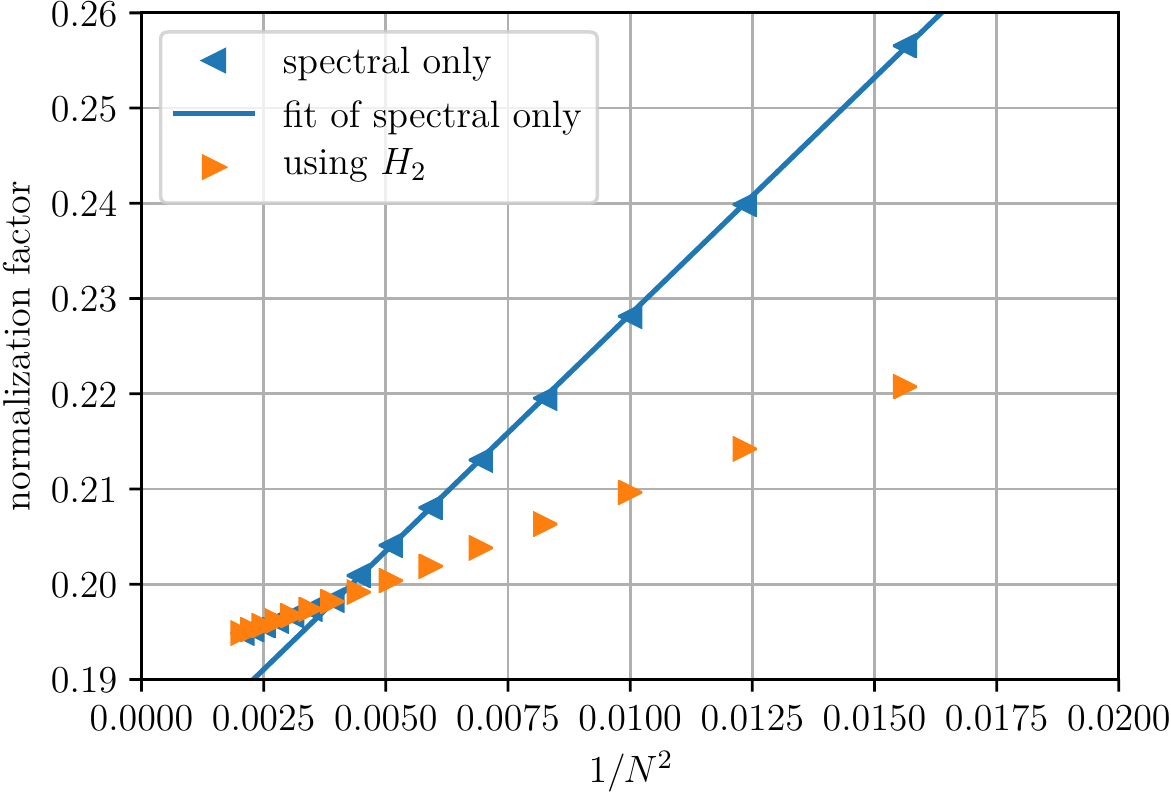}
\caption{\label{fig:ANNNI_normfacs} ANNNI model lattice normalization factors from the spectrum only (assuming $|T\rangle$ is the lowest-energy state with $S=2$) versus using $H_2$ to identify $|T\rangle$. These differ for $N<16$ due to finite-size corrections which shift the energy of another state with $S=2$ below that of $|T\rangle$. See Fig.~\ref{fig:ANNNI_pri}. We fit the spectral data for $N=8\dots 15$ to illustrate the large error made when $|T\rangle$ is incorrectly identified.
}
\end{figure}

\begin{figure}
\includegraphics[width=0.9\linewidth]{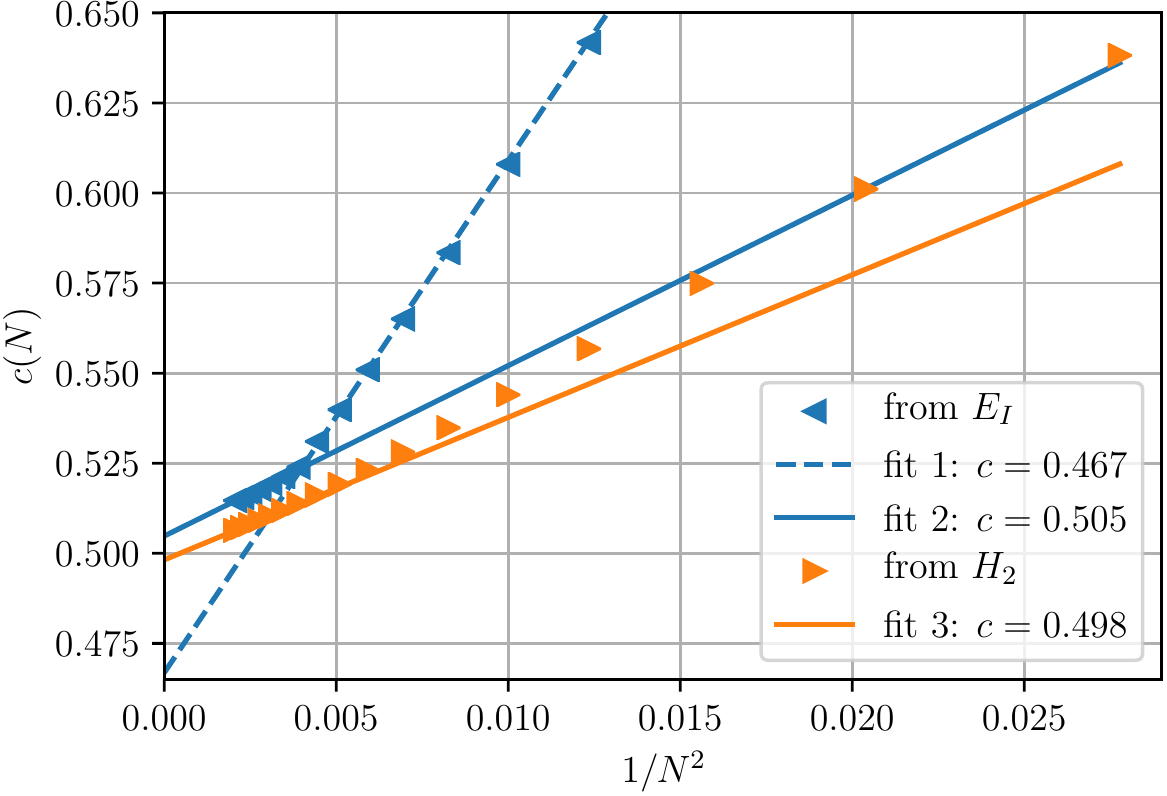}
\caption{\label{fig:ANNNI_c} The central charge for the ANNNI model, comparing estimates using $H_2$ according to~\eqref{eq:c_lat2} with estimates obtained from the ground-state energy $E_{\eye}$ using~\eqref{eq:FS_spec} (after subtracting the extrapolated extensive contribution) \cite{\Henkel}. The sudden change in slope of the $E_{\eye}$ data points is due to erroneous normalization for $N<16$: See Fig.~\ref{fig:ANNNI_normfacs}. Extrapolation is performed using linear regression. We fit the $E_I$ data for $N=8\dots 15$ in fit 1 to illustrate the effects of incorrect normalization. For comparison, in fits 2 and 3, we use $N=17\dots 22$. The CFT value is $c=\frac{1}{2}$. The x-axis is chosen to be $N^{-(4-2)}$ to match the leading finite-size correction to the energy, which is due to an operator with $\Delta=4$, as in the Ising model.
}
\end{figure}

Unlike in the Potts model, the observed corrections to $H^\ANNNI_n$ can only come from perturbation of $h^ \CFT(x)$ by irrelevant operators from the \emph{identity} conformal tower, since operators from any other conformal tower would lead to tower mixing. Furthermore, there must be a perturbation that is not present in the Ising model, which did not exhibit the corrections we see here. One allowed identity-tower perturbation of $h^\CFT(x)$ is the field operator $T\overline{T}(x)$, corresponding to the state $L^\CFT_{-2}\overline{L}{}^\CFT_{-2}|\eye\rangle$, which is suppressed in the Ising model~\cite{reinicke_analytical_1987}, but is allowed in general. It seems a likely candidate to cause the observed correction to $H^\ANNNI_1 |\eye\rangle$ since adding it to $h^\CFT(x)$ in~\eqref{eq:Hn_L_CFT} for $n=1$ would result in the usual $H^\CFT_1$ plus a Fourier mode of $T\overline{T}(x)$ which, applied to $|\eye\rangle$, would produce global descendant states of the (quasiprimary) state $L^\CFT_{-2}\overline{L}{}^\CFT_{-2}|\eye\rangle$, including one at level $\Delta=5$. Indeed, in \cite{zou_conformal_2017} we confirm that this perturbation is present in the ANNNI model.

Regarding finite-size corrections to the energies, we note that they are severe enough so that, at $N=14$, the states $|T\rangle$ and $|\overline{T}\rangle$ are \emph{not} the lowest-energy states with $|S|=2$, as is often assumed when normalizing the Hamiltonian density (see Sect.~\ref{subsec:normalization}). Where this occurs, identifying $|T\rangle$ using $H^\ANNNI_2$ is clearly advantageous. Indeed, we observe in Fig.~\ref{fig:ANNNI_normfacs} that the difference in the normalization factors obtained is significant for affected system sizes.

Finally, in Fig.~\ref{fig:ANNNI_c} we demonstrate that the central charge estimated using~\eqref{eq:c_lat2} remains accurate away from integrability. Furthermore, we compare the estimate to that obtained from the scaling of the ground state energy $E_\eye$ \cite{\Henkel}, finding the estimates to be comparable as long as the Hamiltonian is properly normalized, which requires the use of $H^\ANNNI_2$ at small system sizes.

\section{Discussion}

In this paper we have proposed and demonstrated automated procedures for extracting conformal data from generic local quantum spin chains using the Hamiltonian density Fourier modes~$H_n$, first introduced as lattice representations of conformal generators by Koo and Saleur~\cite{koo_representations_1994}. In particular, we explained how to use the~$H_n$ to systematically identify the lattice energy eigenstates corresponding to Virasoro primary and quasiprimary operators of the CFT, as well as how to assign the remaining eigenstates to conformal towers. Furthermore, our demonstration included a nonintegrable model (the ANNNI model), confirming that the so-called Koo-Saleur formula continues to behave as expected away from integrability.

To extract accurate conformal data, one must examine systems of sufficient size, such that non-universal finite-size corrections (e.g.\ due to irrelevant perturbations) are manageable. This is often impossible using exact diagonalization techniques, which we applied to obtain spectra and low-energy eigenstates for the present work, since the computational cost scales exponentially in the system size. Fortunately, our proposals for extracting conformal data using the Hamiltonian density Fourier modes $H_n$ are independent of the method used to diagonalize $H$ and can also be implemented using more sophisticated tools, such as periodic matrix product states, allowing the analysis of critical quantum spin chains with hundreds of spins~\cite{zou_conformal_2017}.

This work contributes toward the overarching goal of, given a generic critical quantum spin chain Hamiltonian $H$, determining the conformal data that specifies the emergent CFT. Indeed, the identification of the Virasoro primary states within the low-energy spectrum is an essential part of this task, one that cannot be accomplished in general using only the spectral information in~\eqref{eq:FS_spec}, but which is made possible by using the lattice operators $H_n$. In order to complete this long-standing research program, a systematic way of determining the OPE coefficients relating the primary operators to each other is still missing (although progress can be made in particular cases -- see for example \cite{read_enlarged_2007, read_associative-algebraic_2007, gainutdinov_lattice_2013}). As it turns out, however, the methods discussed in this paper can be combined with other techniques in order to also estimate the OPE coefficients on the lattice \cite{zou_upcoming_2017}.

Finally, we remark that the action of lattice Virasoro generators in the low-energy subspace of quantum spin chains has found applications beyond the extraction of conformal data. For example, these techniques are used in~\cite{zou_conformal_2017} to study the RG flow between two CFTs, and in~\cite{milsted_upcoming_2017}
 to attach a geometric meaning to tensor networks that discretize a path integral.

\begin{acknowledgments}
We are grateful to John Cardy, Qi Hu, Vaughan Jones, Tobias J. Osborne, Frank Verstraete, Yuan Wan and Yijian Zou for helpful and stimulating discussions. We also thank Jerôme Dubail, Hosho Katsura, Hubert Michel Saleur, and Romain Vasseur for kindly pointing out Ref.~\cite{koo_representations_1994}, of which we were not aware while preparing the first version of this paper.  Finally, we thank the referees who reviewed this paper for providing useful comments that led to improvements in our presentation. The authors acknowledge financial support from the Simons Foundation (Many Electron Collaboration). This research was supported in part by Calcul Québec and Compute Canada, as well as by Perimeter Institute for Theoretical Physics. Research at Perimeter Institute is supported by the Government of Canada through the Department of Innovation, Science and Economic Development Canada and by the Province of Ontario through the Ministry of Research, Innovation and Science.
\end{acknowledgments}

\appendix

\section{Lattice momentum density}
\label{app:mom}

The Virasoro algebra \eqref{eq:Virasoro} fulfilled by the operators \eqref{eq:LCFT} together with~\eqref{eq:Hn_L_CFT}, implies
\begin{align}
  [H^\CFT_n, H^\CFT_m] = (n-m)(L^\CFT_{n+m} - \overline{L}^\CFT_{-(n+m)})
\end{align}
so that we may construct lattice analogues of $L^\CFT_{n}$ and $\overline{L}^\CFT_{m}$ as \cite{koo_representations_1994}
\begin{align}
  L_n &\equiv \frac{1}{2}(H_{+n} + \frac{1}{n}[H_{+n},H_0]) \\
  \overline{L}_n &\equiv \frac{1}{2}(H_{-n} + \frac{1}{n}[H_{-n},H_0]).
\end{align}
This is equivalent to defining a momentum density
\begin{align}
  p_j \equiv \ic[h_j, h_{j-1}]
\end{align}
which satisfies the lattice energy-momentum conservation law
\begin{align}
  \partial_t h_j = \ic [H,h_j] = p_{j+1} - p_j,
\end{align}
and constructing $L_n$ and $\overline{L}_m$ as
\begin{align} 
  L_n \equiv \frac{N}{2\pi} \sum_{j=1}^N e^{+\ic j n \frac{2\pi}{N}} T_j, \qquad
  \overline{L}_n \equiv \frac{N}{2\pi} \sum_{j=1}^N e^{-\ic j n \frac{2\pi}{N}} \overline{T}_j,
\end{align}
with
\begin{align}
  T_j \equiv \frac{1}{2}(h_j + p_j), \qquad \overline{T}_j \equiv \frac{1}{2}(h_j - p_j),
\end{align}
in analogy with the CFT definition of the Virasoro generators~\eqref{eq:LCFT}.

We find in practice that $L_n$ and $\overline{L}_m$ defined for the Ising model have more severe finite-size corrections than $H^\TFI_n$ (see Sect.~\ref{sec:res_ising}). In particular, they connect states with the wrong descendants, although they still do not mix conformal towers.

There is an obvious reason for these additional corrections, which come from finite-size corrections to the energy. Consider the action of $L_n$ on an energy eigenstate $|\Delta\rangle$ of a lattice Hamiltonian $H$. We first assume that $H_n|\Delta\rangle = a|\Delta \!-n\rangle + b|\Delta\!+n\rangle$ such that
\begin{align}
  H_0 |\Delta \rangle &= (\Delta + \epsilon)|\Delta\rangle, \\
  H_0 |\Delta\! - n\rangle &= (\Delta - n + \epsilon')|\Delta\! - n\rangle, \\
  H_0 |\Delta\! + n\rangle &= (\Delta + n + \epsilon'')|\Delta\! + n\rangle,
\end{align}
where $\epsilon$, $\epsilon'$, $\epsilon''$ represent finite-size corrections to the energy, which will generally be different for each energy eigenstate. This scenario is consistent with $a|\Delta - n\rangle$ and $b|\Delta + n\rangle$ being the lattice counterparts of the CFT states $L^\CFT_n|\Delta\rangle^\CFT$ and $\overline{L}^\CFT_{-n}|\Delta\rangle^\CFT$, respectively.
We then find
\begin{align}
  2L_n|\alpha\rangle = & \left(1 + \frac{\Delta + \epsilon}{n} \right) (a|\Delta\!-n\rangle + b|\Delta\!+n\rangle) \\ &- \left( \frac{\Delta + \epsilon'}{n} - 1\right) a |\Delta\!-n\rangle \\ &- \left( \frac{\Delta + \epsilon''}{n} + 1 \right) b |\Delta\!+n\rangle,
\end{align}
where in case $\epsilon=\epsilon'=\epsilon''$ almost all terms cancel and we are left with
\begin{align}
  L_n|\Delta\rangle = a|\Delta\!-n\rangle,
\end{align}
as expected. As noted above, however, generally $\epsilon \neq \epsilon' \neq \epsilon''$ and the cancellation is prevented, leading to an erroneous matrix element of $L_n$ connecting $|\Delta\rangle$ and $|\Delta\!+n\rangle$.

\section{Degeneracies and quasiprimary states}
\label{app:degenerate_qp}

In Figs.~\ref{fig:ising_global_towers} and~\ref{fig:ANNNI_global_towers} we plot the spectra of the Ising and ANNNI models, respectively, at size $N=14$, showing global conformal towers of the quasiprimary states $|T\rangle$ and $|\overline{T}\rangle$. We find in both cases that a linear combination $|\varphi_Q\rangle \equiv a|\varphi_1\rangle + b|\varphi_2\rangle$ of lattice energy eigenstates $|\varphi_1\rangle, |\varphi_2\rangle$ belonging to the (Virasoro) conformal tower of $\eye$ at level $\Delta \approx 4$, $S=4$, fulfills the quasiprimary condition~\eqref{eq:QPstates_Hn} to numerical precision:
\begin{align} \label{eq:qp_ising_lat}
  \Gamma_{\varphi_Q} (H_1 + H_{-1}) |\varphi_Q\rangle \approx 0,
\end{align}
where $\Gamma_{\varphi_Q}$ projects onto states with energy lower than the energy expectation value of $|\varphi_Q\rangle$.
The situation is analogous for the $|\overline{T}\rangle$ descendants.

In the CFT, where the states of the $I$ conformal tower at $\Delta=4$, $S=4$ are degenerate in energy and momentum (see Fig.~\ref{fig:ising_towers}), there is also a quasiprimary state in the corresponding degenerate subspace. We wish to confirm that the lattice state $|\varphi_Q\rangle$ corresponds to the quasiprimary in the CFT. First, we note that, from~\eqref{eq:Virasoro} and~\eqref{eq:QPstates}, the CFT quasiprimary may be built as
\begin{align}
  |\varphi^\CFT_Q\rangle \propto \left( (H^\CFT_{-1})^2 - \frac{4\Delta_T + 2}{3} H^\CFT_{-2} \right)|T\rangle,
\end{align}
which can be seen to be annihilated by $L^\CFT_{1}$. We may construct an analogous state on the lattice as
\begin{align}
  |\tilde \varphi_Q\rangle \propto \left((H_{-1})^2 - \frac{4\Delta_T + 2}{3} H_{-2} \right)|T\rangle.
\end{align}
Doing so we find that, to high precision,
\begin{align}
  \Gamma_{\varphi_Q} (H_1 + H_{-1}) |\tilde \varphi_Q\rangle \approx 0,
\end{align}
and that furthermore $|\tilde \varphi_Q\rangle$ is approximately equal to $|\varphi_Q\rangle$ of~\eqref{eq:qp_ising_lat}, with appropriate normalization. This confirms that the criterion~\eqref{eq:QPstates_Hn} for quasiprimary states on the lattice correctly distinguishes linear combinations of lattice eigenstates that correspond to CFT quasiprimary operators.

We remark here on the observation that degenerate quasiprimary and global secondary states are \emph{mixed} by finite-size corrections to the energy (even when Virasoro conformal towers are not mixed) so that the quasiprimary lattice state is formed by a linear combination of energy eigenstates with \emph{different} energies (which become degenerate in the limit $N \rightarrow \infty$). In the presence of finite-size effects that mix \emph{Virasoro} conformal towers, it could also happen that primary states are mixed with Virasoro descendant states, although we did not observe this in the models tested in this work.

\end{document}